\documentclass[10pt,twocolumn,letterpaper]{article}

\usepackage[pagenumbers]{cvpr} % To force page numbers, e.g. for an arXiv version

\usepackage{makecell}

\definecolor{cvprblue}{rgb}{0.21,0.49,0.74}
\usepackage[pagebackref,breaklinks,colorlinks,allcolors=cvprblue]{hyperref}
\usepackage{multirow}
\def\paperID{} % *** Enter the Paper ID here
\def\confName{CVPR}
\def\confYear{2026}

\newcommand{\mname}{{PartMotionEdit}}
\title{\mname: Fine-Grained Text-Driven 3D Human Motion Editing \\ via Part-Level Modulation}

\author{Yujie Yang$^{1}$, \, Zhichao Zhang$^{1}$, \, Jiazhou Chen$^{1,2,*}$, \, Zichao Wu$^{3}$ \\
$^{1}$Zhejiang University of Technology \\ 
$^{2}$Zhejiang Key Laboratory of Visual Information Intelligent Processing \\ 
$^{3}$Hangzhou Dianzi University \\
$^*$ Corresponding author: cjz@zjut.edu.cn
}

\begin{document}
\maketitle
\begin{abstract}

Existing text-driven 3D human motion editing methods have demonstrated significant progress, but are still difficult to precisely control over detailed, part-specific motions due to their global modeling nature. In this paper, we propose PartMotionEdit, a novel fine-grained motion editing framework that operates via part-level semantic modulation. The core of PartMotionEdit is a Part-aware Motion Modulation (PMM) module, which builds upon a predefined five-part body decomposition. PMM dynamically predicts time-varying modulation weights for each body part, enabling precise and interpretable editing of local motions. To guide the training of PMM, we also introduce a part-level similarity curve supervision mechanism enhanced with dual-layer normalization. This mechanism assists PMM in learning semantically consistent and editable distributions across all body parts. Furthermore, we design a Bidirectional Motion Interaction (BMI) module. It leverages bidirectional cross-modal attention to achieve more accurate semantic alignment between textual instructions and motion semantics. Extensive quantitative and qualitative evaluations on a well-known benchmark demonstrate that PartMotionEdit outperforms the state-of-the-art methods.

\end{abstract}    
\section{Introduction}
\label{sec:intro}

Human motion generation serves a critical role in diverse applications, including animation, virtual and augmented reality (VR/AR), robotics, and sports analysis.
In recent years, with advances in deep learning, text-driven 3D human motion generation has achieved substantial progress.
Leveraging large-scale text-motion paired datasets (e.g., HumanML3D~\cite{guo2022generating}, BABEL~\cite{punnakkal2021babel}, KIT-ML~\cite{plappert2016kit}, etc.), diffusion models exhibit robust generative capabilities in this field, enabling the gradual generation of high-quality human motions from textual descriptions starting from random noise, e.g., MDM~\cite{tevet2023human}, MotionDiffuse~\cite{zhang2024motiondiffuse}, TMR~\cite{petrovich2023tmr}, etc.

Despite the remarkable success of motion generation techniques, single-step generation of satisfactory motions remains a non-trivial and unresolved challenge, particularly for fine-grained motions.
Users often need to modify prompts and perform iterative generations repeatedly to obtain a desired outcome, which is time-consuming and labor-intensive.
To this end, researchers have instead turned their attention to motion editing: a task that entails modifying an existing motion sequence via natural language instructions to yield motions aligning with fine-grained demands.

Early motion editing methods primarily relied on manual specification of body parts or temporal segments, employed techniques like motion retargeting, style transfer or motion interpolation, to achieve local modifications.
Text-based motion editing methods have emerged in recent years~\cite{goel2024iterative,karunratanakul2023guided,zou2024parco}, but most of them are typically limited to specific edit types (e.g., adjustments requiring manual part selection or operations limited to adding/removing sub-actions), due to the lack of large-scale open-source benchmark.
To address this limitation, MotionFix~\cite{athanasiou2024motionfix} curated a large-scale triplet dataset dedicated to human motion editing and proposed an instruction-driven framework for the task, with models trained on this dataset able to preserve overall motion continuity.

SimMotionEdit~\cite{li2025simmotionedit}, building upon MotionFix, proposes a novel auxiliary task for motion similarity prediction, explicitly predicting the frame-wise similarity curve between source and target motions.
%This compels the model to first grasp when and to what degree the instruction induces motion changes prior to executing the edit.
This design improves the spatiotemporal accuracy and overall coherence of edits.
However, the predicted similarity relies on full-body poses and lacks decomposition to the granularity of individual joints or body parts.
Consequently, when an editing instruction involves multiple independent parts demanding differentiated handling, the model’s control capability is inadequate.
%Additionally, the interaction between textual and motion features is relatively coarse-grained, lacking explicit alignment between body parts and the fine-grained semantics of linguistic instructions, resulting in imprecise understanding and execution of complex instructions. [这不是主要局限，这里先不讲了]

In this paper, we propose PartMotionEdit, a novel fine-grained text-driven 3D human motion editing framework.
To improve local control capability, we propose a Part-aware Motion Modulation (PMM) module. This module divides the human body into five key parts and dynamically predicts the editing weights of each body part across timesteps.
It thus helps to intelligently decide \emph{where to edit} and \emph{how to edit}.
To guide the PMM module in learning a semantically consistent part editing distribution, we further introduce a Part-level similarity curve Supervision Mechanism (PSM). It imposes dual-layer normalization constraints on the spatial displacement and rotation differences of each body part respectively, to generate part-level similarity curves that provide precise supervisory signals for the PMM module.
To address the insufficient interaction between textual and motion features, we additionally propose a Bidirectional Motion Interaction (BMI) module. It adopts a bidirectional cross-modal attention mechanism to enable more precise semantic alignment, thus ensuring edited motions are highly consistent with textual intent.

We summarize our core contributions as follows:

\begin{itemize}
\item \mname, a text-driven 3D human motion editing framework with a Bidirectional Motion Interaction (BMI) module to effectively improve the consistency between text instructions and generated motions.

\item A Part-aware Motion Modulation (PMM) module, which decomposes the human body into 5 parts and achieves fine-grained controls by automatically predicting temporal part-wise weights with a multi-part semantic similarity curve supervision.

\item Experiments on the MotionFix dataset~\cite{athanasiou2024motionfix} show that \mname\ outperforms SOTA methods in key metrics.
\end{itemize}

\section{Related Works}
\label{sec:related}

\begin{figure*}[!ht]
\centering
    \includegraphics[width=\linewidth]{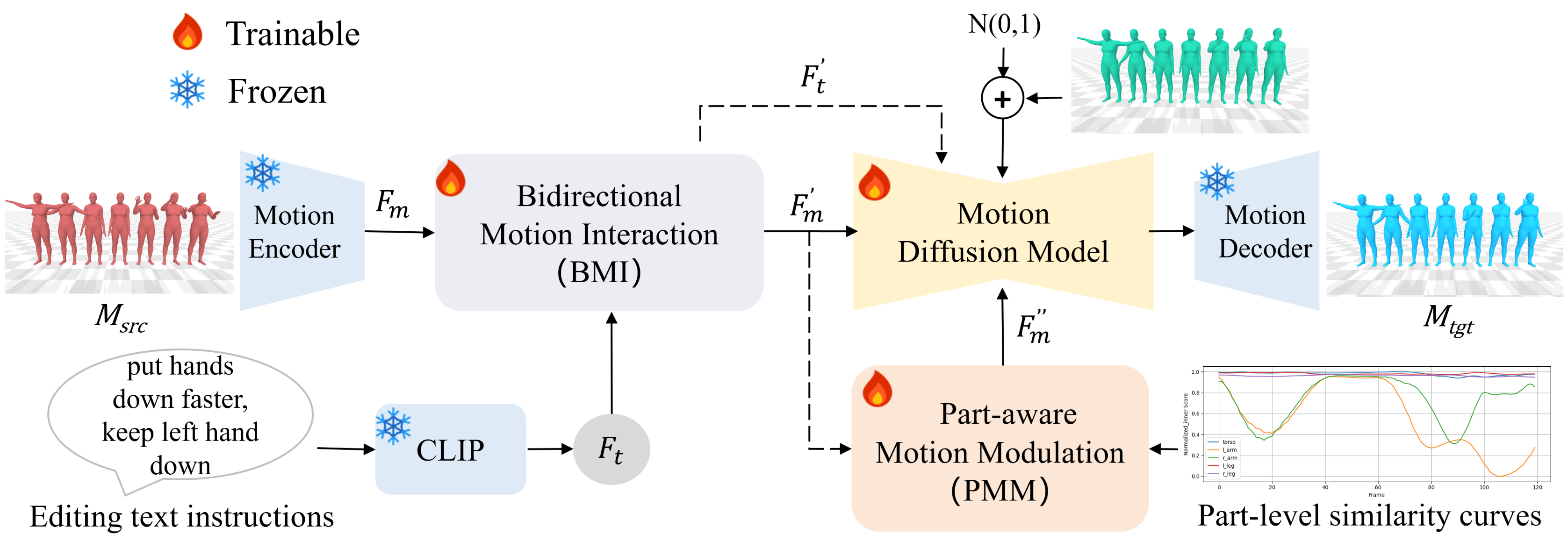}
    \caption{\noindent \textbf{Overview of \mname.} It generates a target motion that modifies the given source motion according to the specified text instructions. It has three trainable core modules: the Bidirectional Motion Interaction (BMI) module, the motion diffusion model, and the Part-aware Motion Modulation (PMM) module with a multi-part similarity curve supervision, and three frozen modules (i.e., the motion encoder, the motion decoder and the CLIP model).
        \label{fig:overview}
        \vspace{0.5em}
    }
\end{figure*}

%-------------------------------------------------------------------------
\subsection{Text-based motion generation}
%文本驱动的 3D 人体动作生成是动作编辑的技术基础，其核心目标是将自然语言描述转化为连贯、真实的人体动作序列，已在动画制作、人机交互等领域展现出重要价值。早期方法多依赖变分自编码器（VAE）等生成模型，通过动作类别标签进行条件约束，但仅支持有限的动作类型生成，难以应对自由文本输入场景 [1,2]。随着 HumanML3D、BABEL 等大规模文本 - 动作配对数据集的发布 [3,4]，基于扩散模型的生成方法成为主流，显著提升了动作的真实性和文本对齐性。其中，人体动作扩散模型（MDM）[5] 首次将扩散模型应用于该任务，通过文本特征引导噪声迭代 denoising 生成动作；MotionDiffuse [6] 进一步优化了文本 - 动作的语义映射，实现了更精细的动作生成。
%为提升生成可控性，研究者们引入了时空组合机制，支持多动作序列的时序拼接 [7] 和多动作的空间并行生成 [8]；部分方法还实现了关节级别的精准控制，如 OmniControl [9] 支持任意关节在任意时刻的运动调节，MoMask [10] 通过掩码策略增强局部动作生成的准确性。此外，文本编码方案的优化也推动了性能提升，主流方法均采用 CLIP 模型 [11,12] 提取文本特征，实现跨模态语义的有效对齐。
Text-driven 3D human motion generation~\cite{zhu2023human} serves as the technical foundation for motion editing, with its core objective being to convert natural language descriptions into coherent and realistic human motion sequences. 
%It has demonstrated significant value in fields such as animation production and human-computer interaction. 
Early approaches mostly relied on generative models based on Variational Autoencoders (VAEs)~\cite{yan2018mt,ling2020character}, which applied conditional constraints through motion category labels. However, these VAE-based methods only support the generation of a limited range of motion types and struggle to handle scenarios involving free-text inputs~\cite{petrovich2021action,bie2022hit}.

With the release of large-scale text-motion paired datasets such as HumanML3D and BABEL~\cite{guo2022generating,punnakkal2021babel}, generative methods based on diffusion models become the mainstream, significantly enhancing the realism of motions and the alignment between motion and text. The Motion Diffusion Model (MDM)~\cite{tevet2023human} is a pioneer work to apply diffusion models to motion generation through a text feature-guided iterative noise denoising process. And MotionDiffuse~\cite{zhang2024motiondiffuse} further optimized the semantic mapping between text and motion, enabling more refined motion generation.

To improve the controllability of generation, researchers have introduced spatiotemporal composition mechanisms, which support the temporal stitching of multiple motion sequences~\cite{athanasiou2022teach} and the spatial parallel generation of multiple motions~\cite{athanasiou2023sinc}. Some methods have also achieved precise joint-level control: for instance, OmniControl supports the adjustment of the movement of any joint at any time~\cite{xie2024omnicontrol}, while MoMask enhances the accuracy of local motion generation through a masking strategy~\cite{guo2024momask}. In addition, the optimization of text encoding schemes has also driven performance improvements. Mainstream methods all adopt the CLIP model~\cite{tevet2022motionclip,maldonado2025moclip} to extract text features, achieving effective cross-modal semantic alignment.

%-------------------------------------------------------------------------
\subsection{Text-based motion editing}
3D human motion generation has achieved impressive development. However, it is still difficult to satisfy complex or high-requirement tasks in a single step of motion generation, further editing is often inevitable. Early motion editing methods relied mostly on handcrafted constraints or rules:
%-based design, requiring users to specify explicit conditions like spatial ranges or body parts
spatial-temporal constraint-based methods optimize trajectories for simple adjustments~\cite{gleicher1997motion,gleicher2001motion,starke2020local}; style transfer methods transfer motion style but lack fine-grained semantic editing~\cite{yin2023dance}; pose editing methods focus on static correction and struggle with dynamic temporal coherence~\cite{oreshkin2021protores}. However, none of them supports free-form editing based on natural language.

In recent years, text-driven automatic motion editing has become a research hotspot~\cite{zhang2025text}. For instance, PoseFix realized text-based static pose correction~\cite{delmas2023posefix}, MotionCLR achieved training-free editing via attention manipulation~\cite{motionclr}, CoMo~\cite{huang2024controllable} and FineMoGen~\cite{zhang2023finemogen} used large language models for editing instructions. However, these methods struggled to balance source motion preservation and text alignment due to the lack of training data. 

To this end, Athanasiou et al. proposed MotionFix\cite{athanasiou2024motionfix}, the first text-driven dynamic motion editing dataset with 6,730 ``source motion - target motion - editing text'' triplets~\cite{athanasiou2024motionfix}. They designed the conditional diffusion model, enabling dual-conditional editing of source motion and text, outperforming MDM-based baselines. 

Based on MotionFix, SimMotionEdit~\cite{li2025simmotionedit} introduced an auxiliary motion similarity prediction task to locate key editing frames. Motion editing and motion similarity prediction are then jointly trained in a multi-task paradigm to foster the learning of semantically meaningful representations.
Due to its global modeling nature, this method often produces inaccurate local edits and causes disturbances in non-target parts, as it tends to fail to capture fine-grained inter-part differences. To this end, we decompose the human body into 5 parts as Local Motion Phase~\cite{starke2020local}, and further design a multi-part similarity curve supervision mechanism to improve the accuracy for each part.

%-------------------------------------------------------------------------
\subsection{Fine-grained local editing}
% local motion phase - dynamic motion blending
Prior work has explored methods for local motion editing. For example, Local Motion Phase~\cite{starke2020local} decomposes a full-body motion into locally synchronized phases to enable precise and temporally consistent part-level adjustments, and MotionRefit~\cite{jiang2025dynamic} employs Gaussian-field pose fitting with predefined joint masks to achieve text-guided local editing. Unlike these methods, our approach learns part-level editability directly from motion-text alignment. This capability enables adaptive local modifications without relying on predefined mask annotations or explicit target poses.

Fine-grained local editing has been extensively studied in image editing, with notable examples including InstructPix2Pix~\cite{brooks2023instructpix2pix}, VAREdit~\cite{mao2025visual}, and UniWorld-V2~\cite{li2025uniworld}. Our work is inspired by a recent text-based image editing approach introduced by Lin et al.~\cite{lin2024text}. We adapt their concept of \emph{learnable regions} into the notion of \emph{learnable parts}, and propose a Part-aware Motion Modulation module to adaptively assign weights to different body parts. By integrating a multi-part semantic similarity curve supervision mechanism, our method achieves high precision and locally controllable motion editing.

%Text-driven 3D human motion editing requires fine-grained control while preserving global coherence, but existing methods either need manual body part/keyframe specification or miss inter-part subtleties—disturbing non-target regions. This issue has been well studied in the field of image editing, where many local image editing methods follow a pipeline ``automatic localization - precise modification - global coherence'', such as InstructPix2Pix~\cite{brooks2023instructpix2pix}, VAREdit~\cite{mao2025visual}, and UniWorld-V2~\cite{li2025uniworld}. 

%Our work is inspired by the mask-free region-based editing approach proposed by Lin et al.~\cite{lin2024text}. This method leverages an existing pre-trained text-to-image model and introduces a bounding box generator to identify the editing regions that are aligned with the textual prompts. In this paper, the idea of \emph{learnable regions} is extended to \emph{learnable parts}. We propose a Part-aware Motion Modulation module to predict weights for different body parts with a multi-part semantic similarity curve supervision, achieving high accuracy and local control of motion editing.

\section{Methodology} 
\label{sec:method}

%-------------------------------------------------------------------------
\subsection{Overview}

%给定文本编辑指令P与源动作序列M_{src}，我们的目标是生成与指令语义一致的目标人体三维动作序列M_{tgt}。其中动作序列M\in\R^{T×D}，序列帧数为T，每个运动帧的维度D=207，由3个全局平移特征、12个全局方向特征以及192个身体姿势特征组成。
%Given the text editing instruction $P$ and the source motion sequence $M_{src}$ , our goal is to generate a target 3D human motion sequence $M_{tgt}$ that is semantically consistent with the instruction. Here, the motion sequence $M\in\mathbb{R}^{T×D}$ , where $T$ denotes the number of frames in the sequence, and $D = 207$ represents the dimension of each motion frame. This dimension $D$ consists of 3 global translation features, 12 global orientation features, and 192 body pose features.
Given a text editing instruction $P$ and a source motion sequence $M_{src}$, our objective is to produce a target 3D human motion sequence $M_{tgt}$ that semantically aligns with this instruction. Each motion sequence is represented as $M \in \mathbb{R}^{T \times D}$. $T$ is the number of sequence frames, $D = 207$ is the feature dimension for each frame, 3 for global translation, 12 for global orientation, and 192 for body poses.

%我们在图X中呈现了我们的网络框架内容，首先将源动作序列M_{src}经过预训练的动作编码器，提取动作特征F_{m}；同时，将文本编辑指令P输入预训练CLIP文本编辑器，获得文本特征F_{t}。随后，我们将F_{m}和F_{t}一同输入至双向语义交互模块（BMI，详见Sec.x.x），以实现跨模态特征的双向语义对齐，输出语义一致的文本特征F_{t}'和动作特征F_{m}'。
%We present the content of our network framework in Figure~\ref{fig:overview}. First, the source motion sequence $M_{src}$ is fed into a pre-trained motion encoder to extract motion features $F_{m}$; meanwhile, the text editing instruction $P$ is input into a pre-trained CLIP text encoder to obtain text features $F_{t}$. Subsequently, we feed $F_{m}$ and $F_{t}$ together into the Mutual Semantic Interaction Module (BMI, Sec. 3.3) to achieve bidirectional semantic alignment of cross-modal features, and output semantically consistent text features $F_{t}'$ and motion features $F_{m}'$.
Figure~\ref{fig:overview} illustrates the overall architecture of our proposed framework. Specifically, the source motion sequence $M_{src}$ is first processed by a pre-trained motion encoder to extract motion features $F_{m}$. Simultaneously, the text editing instruction $P$ is fed into a pre-trained CLIP text encoder to obtain text features $F_{t}$. Subsequently, both $F_{m}$ and $F_{t}$ are passed through a Mutual Semantic Interaction Module (BMI, Sec. \ref{tit.BMI}), which enables bidirectional semantic alignment between the cross-modal features and outputs semantically aligned text features $F_{t}'$ and motion features $F_{m}'$.

%接着，动作特征F_{m}'将会被送入部位感知调制模块（PMM，详见Sec.x.x），并辅以部位级相似度曲线的监督信号（Sec.x.x）用于预测各身体部位的编辑权重，从而对动作特征进行部位级的精细调制，得到调制后的特征F_{m}''。最后，Motion Diffusion Model以F_{m}''为输入并结合噪声序列与文本特征F_{t}'作为条件，在时序潜空间中逐步去噪，生成符合文本语义的编辑动作序列。
%Next, the motion feature $F_{m}'$ is fed into the Part-Aware Modulation Module (PMM, Sec. 3.4), and is supplemented with the supervision signal from the part-level similarity curve (Sec. 3.2.1) to predict the editing weights of each body part. This enables fine-grained part-level modulation of the motion feature, resulting in the modulated feature $F_{m}''$. Finally, the Motion Diffusion Model takes $F_{m} ''$ as input, and incorporates the noise sequence and text feature $F_{t}'$ as conditions. It gradually denoises in the temporal latent space to generate an edited motion sequence that conforms to the text semantics.
Guided by the Part-level Supervision Mechanism (PSM, Sec. \ref{tit.PSM}), Part-Aware Modulation Module (PMM, Sec. \ref{tit.PMM}) predicts editing weights for each body part. PMM produces fine-grained, part-level modulation of the motion features $F_{m}''$. Finally, the diffusion model takes $F_{m}''$ and $F_{t}'$ as input, generates a motion sequence that is semantically aligned through iterative denoising in the temporal latent space.

%-------------------------------------------------------------------------
\subsection{Part-level Supervision Mechanism (PSM)}\label{tit.PSM}
%在文本驱动的动作编辑任务中，模型需要在理解语言语义的同时，对人体各部位的动作变化进行精确控制。然而现有方法仅在全局尺度上建模源动作与目标动作的整体相似度，未能捕捉到局部身体区域的细粒度语义差异，导致编辑时易出现全局一致而局部失真的现象。为此，本文提出一种部位相似度预测机制，在训练阶段引入部位级相似度监督信号，通过显式回归不同身体部位的时间序列相似度曲线，使模型在生成编辑动作的同时学习身体部位的对应关系与变化趋势。
%In text-driven motion editing tasks, the model needs to accurately control the motion variations of all body parts while understanding the semantic meaning of the input prompts. However, existing methods mainly consider the global similarity between the source and target motions, failing to capture the fine-grained semantic differences in local body parts, resulting in the phenomenon of global consistency but local distortion during editing. To address this issue, we propose a part-level similarity prediction mechanism. During the training stage, it introduces part-level similarity supervision signals and explicitly regresses the similarity curves of time series for different body parts, learning the corresponding relationships and change trends of body parts while generating the expected motions.
In text-driven motion editing tasks, it is crucial for the model to not only comprehend the semantic meaning of the input prompts but also precisely control the motion variations of all body parts. However, existing methods primarily focus on the global similarity between the source and target motions, often overlooking fine-grained semantic differences in local body parts. This limitation leads to edited motions that exhibit global consistency but suffer from local distortions. To address this issue, we propose a part-level similarity prediction mechanism. During training, this mechanism incorporates part-level similarity supervision signals and explicitly regresses the time-series similarity curves for different body parts. This approach enables the model to learn the correspondences and variation trends of body parts while generating the desired motions.

\subsubsection{Part-level similarity curves}
% 首先，我们使用基于人体骨架的语义结构将 22 个关节划分为五个主要身体部位：
To support precise control, we decompose the human skeleton with 22 joints into 5 main body parts (i.e., the torso, left arm, right arm, left leg, and right leg):
\begin{equation}
    G=\{Torse, L_{arm}, R_{arm}, L_{leg}, R_{leg}\}    
\end{equation}

%分别对应躯干、左上肢、右上肢、左下肢与右下肢。为了刻画身体部位在时序维度上的动态差异，我们综合考虑关节的空间位置偏移与局部旋转变化两个因素。对于每个部位gi，我们在时间维度上计算源动作与目标动作在该区域的欧几里得距离作为原始空间位移相似度，即
%They correspond to the torso, left arm, right arm, left leg and right leg respectively. To depict the temporal variation of body parts, we comprehensively consider two factors: the spatial position offset and the local rotation changes of the joints. For each part $g_i\in G$, we calculate the average Euclidean distance between the source motion and the target motion in the temporal dimension as the similarity in the spatial space for this part:
To characterize the temporal variations of each body part, two key factors are taken into account: spatial position offsets and local rotation changes of the joints. For each part $g_i \in G$, we compute the average Euclidean distance in the temporal dimension between the source motion and the target motion, which serves as the spatial similarity measure for that part:

\begin{equation}
    D_{i,t}^{pos} = \frac{1}{|g_i|} \sum_{j \in g_i} \| X_{t,j}^{src} - X_{t,j}^{tgt} \|_2
\end{equation}
\noindent where $X_{t,j}^{src}$/$X_{t,j}^{tgt}$ is 3D position of the $i^{th}$/$j^{th}$ joint in the $t^{th}$ frame, and $|g_i|$ is the number of joints in this body part.
%$D_{i,t}^{pos}$ computes the overall spatial translation distance for each body part.

%此外，仅依赖位置差异难以准确反映姿态层面的变化，因此我们进一步引入局部旋转相似度，用于度量身体部位在关节朝向上的动态一致性。对于每个部位 gi\in G，我们在欧拉角或旋转向量空间中计算对应关节的旋转差异：
%Furthermore, relying solely on spatial translation distances is insufficient to accurately reflect the changes at the posture level. Therefore, we further introduce local rotation similarity to measure the dynamic consistency of body parts along the joint orientation. For each part $g_i \in G$, we calculate the rotation variation of the corresponding joint in the Euler angles or rotation vector space:
Furthermore, spatial translation distances alone are insufficient to accurately capture changes in body posture. To address this, we introduce a local rotation similarity measure to assess the dynamic consistency of body parts in terms of joint orientation. For each part $g_i \in G$, we compute the rotation variation of the corresponding joint in either Euler angles or rotation vector space:
\begin{equation}
    D_{i,t}^{rot} = \frac{1}{|g_i|} \sum_{j \in g_i} \| R_{t,j}^{src} - R_{t,j}^{tgt} \|_2
\end{equation}
\noindent where $R_{t,j}^{src}$/$R_{t,j}^{tgt}$ is 3D position of the $i^{th}$/$j^{th}$ joint in the $t^{th}$ frame. The final similarity in the $t^{th}$ frame is then computed as the weighted sum:
\begin{equation}
    S_{i,t}=-\big( \beta\cdot D_{i,t}^{pos} + (1-\beta)\cdot D_{i,t}^{rot}\big)
\end{equation}
\noindent where the hyperparameter $\beta$ balances spatial translation distance and the rotation variation. The negative sign indicates that a larger similarity value means the motions are more consistent. Simultaneously modeling the variations of translation and rotation can accurately reflect both geometric and motion consistency of body parts.

\subsubsection{Dual-layer normalization}
%Due to the significant differences in scale, speed and amplitude among different motions, directly using the original similarity values as the supervisory signal may lead to instability in the training process. Therefore, we adopt a dual-layer normalization strategy to improve the consistency and comparability of the similarity distribution.

%For each body part $I$, we count the global maximum $S_i^{max}$ and global minimum $S_i^{min}$ similarity of the part within the dataset, and map the similarity to the [0,1] interval to achieve global normalization:
Owing to significant variations in scale, velocity, and amplitude across different motions, directly using raw similarity values as supervision signals may cause training instability. We employ a dual-layer normalization strategy to mitigate this issue. For a body part $i$, we compute the maximum $S_i^{\text{max}}$ and minimum $S_i^{\text{min}}$ of its similarity values across the dataset, and normalize the similarity globally:
\begin{equation}
   \hat{S}_{i,t}=\frac{S_{i,t}-S_i^{min}}{S_i^{max}-S_i^{min}+\epsilon}
\end{equation}
where $S_{i,t}$ denotes the original similarity of the current motion at part $i$ and frame $t$, while $\hat{S}_{i,t}$ represents the similarity after global normalization. This step aims to eliminate the numerical scale differences between different parts, ensuring that the similarity distributions of all parts fall within a unified range.
Subsequently, we calculate the maximum similarity value $\hat{S}_b^{max}$ and the global minimum similarity value $\hat{S}_b^{min}$ across all body parts within a single motion sequence $b$, and perform frame-level normalization to emphasize the relative variation trend within the motion.
\begin{equation}
    \bar{S}_{i,t}=\frac{\hat{S}_{i,t}-\hat{S}_b^{min}}{\hat{S}_b^{max}-\hat{S}_b^{min}+\epsilon}
\end{equation}

%$\bar{S}_{i,t\in b}$ represents the final similarity score curve of motion $b$ at part $i$, where $t$ is the curve parameter. Through this hierarchical normalization process, we can simultaneously maintain cross-part scale consistency and inter-frame relative differences, thereby obtaining more stable and interpretable part-level similarity curves.

%Our dual-layer normalization improves the comparability of similarities among different samples at the global level, while enhances the relative differences between various body parts within a single motion at the local level. This hierarchical normalization strategy enables the model to focus more on the dynamic changes of the edited parts during the subsequent training process, thus effectively avoiding the interference of global scale differences on learning. Finally, we take the part similarity curve $\bar{S}_{i,t}$ of each sample as a supervision signal and input it into the similarity prediction auxiliary head, so as to guide the model to pay attention to the motion differences of the corresponding body regions at different time steps.
\noindent where $\bar{S}_{i,t}$ denotes the final similarity score curve for motion $b$ at part $i$. Through this hierarchical normalization process, we preserve both cross-part scale consistency and inter-frame relative variations, yielding more stable and interpretable part-level similarity curves.

% Our dual-layer normalization enhances the comparability of similarity scores across different samples at the global level, while maintaining relative distinctions among body parts within individual motions at the local level. This hierarchical approach enables the model to focus more effectively on the dynamic variations of edited regions during subsequent training, thereby reducing the influence of global scale discrepancies on the learning process. Finally, we use the part similarity curve $\bar{S}_{i,t}$ of each sample as a supervision signal, which is fed into a similarity prediction auxiliary head to guide the model in capturing motion differences of corresponding body parts across time steps.
Our dual-layer normalization enhances the comparability of similarity scores globally across samples while preserving relative distinctions among body parts locally within each motion. This hierarchical approach allows the model to focus more effectively on the dynamic variations in edited parts, reducing interference from global scale discrepancies. Finally, the part similarity curve $\bar{S}_{i,t}$ of each sample serves as a supervision signal for an auxiliary prediction head, guiding the model to capture temporal motion differences of corresponding body parts.

%-------------------------------------------------------------------------
\subsection{Bidirectional Motion Interaction (BMI)}\label{tit.BMI}
\begin{figure}[!ht]
\centering
    \includegraphics[width=\linewidth]{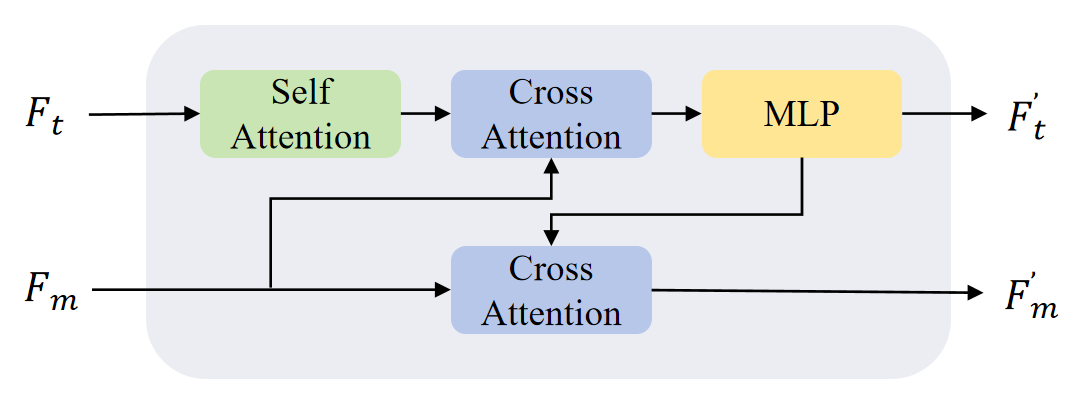}
    \caption{Our Bidirectional Motion Interaction (BMI) module takes textual and motion features ($F_t$ \& $F_m$) as inputs and uses two bidirectional cross attentions to exchange feature information.
        \label{fig:BMI}
        \vspace{0.5em}
        }
\end{figure}

%为了实现文本语义和人体动作特征之间的高效对齐，我们引入了Motion Bidirectional Interaction Module（BMI）。BMI 接收来自文本编码器的文本特征Ft 与来自动作编码器的时序运动特征Fm作为输入，通过双向语义交互机制在语言与动作之间建立细粒度的语义交互，从而提升文本指令与动作片段之间的关联性。经过交互后的输出分别记为 Ft'和 Fm′。
%To achieve efficient alignment between text semantics and human motion features, we introduce the Bidirectional Motion Interaction(BMI). It receives the text feature $F_t$ from the text encoder and the temporal motion feature $F_m$ from the motion encoder as input, and establishes a semantic interaction between prompts and motions through a bidirectional semantic interaction mechanism, enhancing the relevance between text instructions and motion fragments. The output after interaction is recorded as $F'_t$ and $F'_m$ respectively.
To enable effective alignment between textual semantics and human motion features, we propose the Bidirectional Motion Interaction (BMI) module. The module takes as input the text feature $F_t$ from the text encoder and the temporal motion feature $F_m$ from the motion encoder. Through a bidirectional interaction mechanism, it facilitates semantic exchange between prompts and motion sequences, thereby strengthening the relevance between text instructions and motion segments. The resulting outputs after interaction are denoted as $F'_t$ and $F'_m$, respectively.

%如图X所示，它由一个自注意力层（Self-Attention Block）、两个双向交叉注意力层（Bidirectional Cross-Attention Blocks）以及一个多层感知机（MLP）组成。首先对文本特征Ft 进行自注意力建模，以捕获句子内部的语义依赖并增强上下文表达。随后，文本特征 Ft  作为查询，与动作特征Fm通过交叉注意力机制进行交互，其中Fm作为键值（Key/Value）提供时序结构信息进行，从而生成更新后的文本特征表示Ft′ 。接着，得到的 Ft′ 被反向用作键值，而动作特征Fm作为查询再次执行交叉注意力操作与MLP处理，完成动作特征对文本语义的吸收，得到融合后的Fm′。
%As shown in Figure~\ref{fig:BMI}, BMI consists of a Self-Attention Block, two Bidirectional Cross-Attention Blocks, and a Multi-Layer Perceptron (MLP). Firstly, a self-attention model is performed on the text feature $F_t$ to capture the semantic dependencies within the prompt and enhance the context representation. Secondly, the text feature $F_t$ is used as a query to interact with the motion feature $F_m$ through the cross-attention mechanism. Here, $F_m$ serves as the key/value to provide temporal structure information, thereby generating an updated text feature representation $F′_t$. Subsequently, the obtained $F′_t$ is reversely used as the key/value, and the motion feature $F_m$ is used as the query to perform the cross-attention operation and MLP again, completing the absorption of the text semantics by the motion feature and obtaining the fused $F'_m$. 
As illustrated in Figure~\ref{fig:BMI}, the proposed BMI comprises a Self-Attention block, two Bidirectional Cross-Attention blocks, and a Multi-Layer Perceptron (MLP). First, self-attention is applied to the text feature $F_t$ to capture semantic dependencies within the prompt and enhance its contextual representation. Then, $F_t$ serves as the query to interact with the motion feature $F_m$ via cross-attention, where $F_m$ provides temporal structural information as the key and value, yielding an updated text feature $F'_t$. Subsequently, in a reverse interaction, $F'_t$ acts as the key and value while $F_m$ serves as the query. This cross-attention step, followed by an MLP, allows the motion feature to assimilate textual semantics, producing the fused motion feature $F'_m$.

%通过这种双向交互机制，BMI 实现了文本与动作特征的动态融合，使动作表征能够充分感知语言描述中的意图，而语言表示也能捕捉到动作序列中的时序结构。与未使用交互模块或仅采用单向融合的模型相比，BMI 能有效提升语义对齐的精度与动作编辑的可控性。后续章节的实验结果表明，引入 BMI 后的模型在动作一致性与指令响应性方面均取得显著提升。
%Through the bidirectional interaction mechanism, BMI carries out the dynamic fusion of text and motion features, enabling motion representations to fully perceive the intent in the text description, and the text representation can also capture the temporal structure in the motion sequence. Compared with models that do not use interaction modules or only use one-way fusion, BMI can effectively improve the accuracy of semantic alignment and the controllability of motion editing. Experimental results in subsequent chapters show that the models introduced with BMI have achieved significant improvements in both motion consistency and instruction responsiveness.
Through its bidirectional interaction mechanism, BMI facilitates dynamic fusion between text and motion features. This enables motion representations to fully perceive the semantic intent from textual descriptions, while also allowing text representations to capture the temporal structure inherent in motion sequences. Experimental results in later sections demonstrate that models incorporating BMI achieve notable improvements in both motion-text consistency and instruction adherence.
%-------------------------------------------------------------------------
\subsection{Part-aware Motion Modulation (PMM)}\label{tit.PMM}
\begin{figure}[!ht]
\centering
    \includegraphics[width=\linewidth]{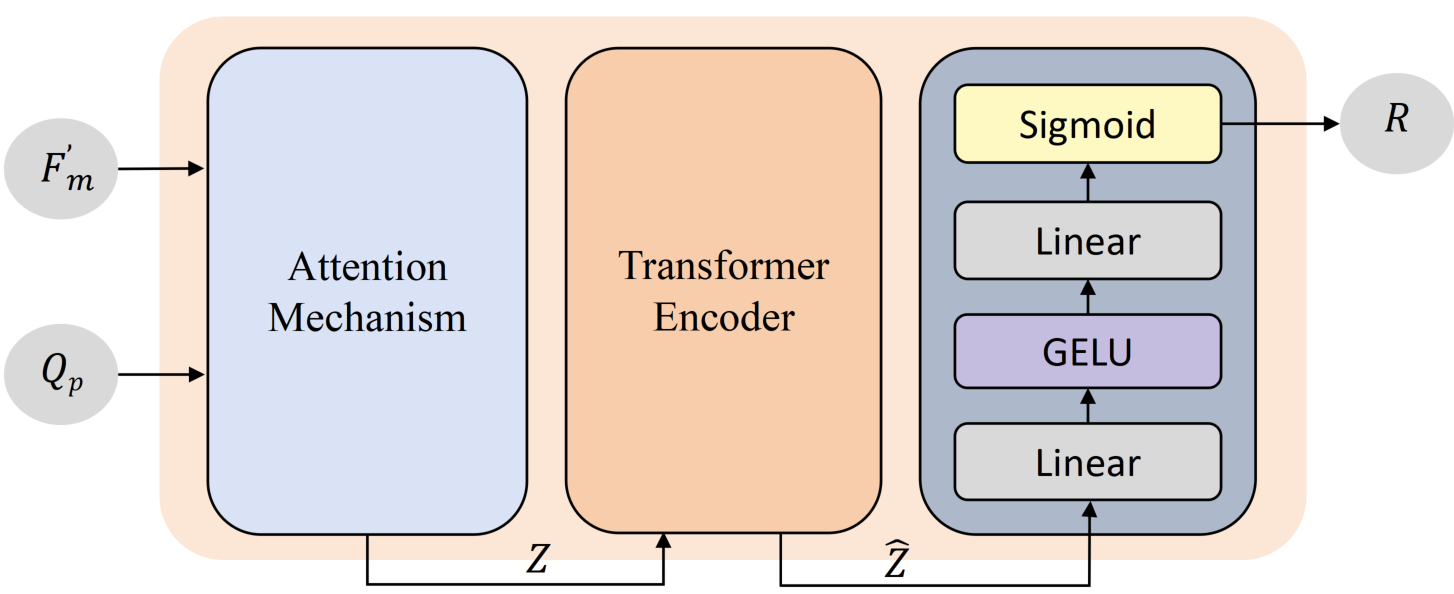}
    \caption{Our Part-aware Motion Modulation (PMM) module processes motion features $F'_m$ with learnable part queries $Q_p$ via attention and transformer, producing part-aware modulation $R$.
        \label{fig:PMM}
        \vspace{0.5em}
        }
\end{figure}

%在经过 BMI 模块的双向语义交互后，我们获得了语义对齐的动作特征 Fm′，其中每个时间步的表示均融合了文本描述的语义信息。然而，仅依赖全局特征进行动作生成，仍难以实现对特定身体区域的精确控制。为此，我们在 Sec. 3.2 中提出了 part-level similarity curve supervision，将全局动作相似度分解为五个主要身体部位的细粒度语义相似度，从而使模型具备初步的部位感知能力。
%After the bidirectional semantic interaction of the BMI module, we obtained the semantically aligned motion feature $F′_m$, in which the features of each time step are fused with the semantic information of the text instructions. However, it is still difficult to achieve high precision using only on global features for motion generation, as shown in Figure~\ref{fig:PMM}. Therefore, we propose the part-level similarity curve supervision in Sec.3.2, which decomposes the global motion similarity into the fine-grained semantic similarity of the five main body parts, so that the model has the preliminary part perception ability.
Following the bidirectional semantic interaction in BMI, we obtain the semantically aligned motion feature $F'_m$, where each time step incorporates semantic information from the text instruction. However, relying solely on global features for motion generation remains challenging for achieving high precision, as illustrated in Figure~\ref{fig:PMM}. To address this, we introduced a part-level similarity curve supervision in Sec. 3.2, which decomposes global motion similarity into fine-grained semantic similarities of the five major body parts. This decomposition equips the model with a preliminary ability to perceive part-level semantics.

%尽管这一机制有效提升了模型对局部差异的敏感性，但在实际动作编辑场景中，不同任务往往仅涉及 1–2 个主要身体部位（如“抬手”主要影响上肢区域，而“下蹲”主要涉及下肢区域），而非全身范围的均匀编辑。由于缺乏显式的部位选择与权重建模，模型仍难以准确判断哪些区域应被重点修改，导致编辑范围模糊、非目标部位受到干扰。
%Although this mechanism effectively increases the model's sensitivity to local differences, in actual motion editing scenarios, different tasks often involve only 1-2 major body parts (e.g., raising hands mainly affects the upper limb area, while squating mainly affects the lower limb area), rather than uniform editing of the whole body. Due to the lack of explicit part selection and weight modeling, the model still has difficulty in accurately determining which areas should be modified with emphasis, resulting in an ambiguous editing scope and interference with non-target areas.
While this mechanism effectively enhances the model's sensitivity to local variations, real-world motion editing tasks typically involve only 1–2 major body parts (e.g., raising arms primarily affects the upper limbs, while squatting mainly influences the lower limbs), rather than requiring uniform editing of the entire body. Due to the absence of explicit part selection, the model struggles to accurately identify which parts should be modified and to what extent, leading to ambiguous editing scopes and unintended interference with non-target areas.

%针对这一问题，我们提出了 Part-aware Motion Modulation（PMM）模块，在动作特征空间中显式预测不同身体部位的可编辑性权重，从而实现更具结构感知与语义一致性的动作编辑。为了让模型具备这种区域敏感性，我们在语义对齐特征 Fm′ 的基础上引入了一组可学习的部位查询向量（learnable part queries）：
%To this end, we propose a Part-aware Motion Modulation (PMM) module for explicitly predicting the regional editable weights of different body parts in the motion feature space, thus enabling more structure-aware motion editing. Firstly, we introduce a set of learnable part queries based on the semantic alignment feature $F′_m$:
To address this challenge, we propose a Part-aware Motion Modulation (PMM) module that explicitly predicts editable part weights for different body parts in the motion feature space, thereby enabling more structurally-aware motion editing. The module first introduces a set of learnable part queries based on the semantically aligned motion feature $F'_m$: $Q_p=\{q_1, q_2,q_3,q_4,q_5\}, q_i\in\mathbb{R}^D$. They correspond to 5 distinct body parts: the torso, $L_{arm}$, $R_{arm}$, $L_{leg}$, and $R_{leg}$ respectively. These vectors are part-aware semantic prototypes. They can adaptively learn characteristic feature patterns of each respective part during training.

%其中每个 qi​ 分别对应躯干（torso）、左臂（l_arm）、右臂（r_arm）、左腿（l_leg）与右腿（r_leg）五个身体部位。这些向量作为区域感知的语义原型，能够在训练过程中自适应地学习各个部位的特征模式。
%These $q_i$ corresponds to five body parts: torso, left arm ($L_{arm}$), right arm ($R_{arm}$), left leg ($L_{leg}$) and right leg ($R_{leg}$). These vectors serve as the semantic prototypes of region-aware and can adaptively learn the feature patterns of various parts during the training process.

%我们为每个身体部位 gi分配一个可学习的部位查询向量 qi，用于捕获该部位在动作序列中的动态相关性。具体而言，模型通过计算区域查询与动作特征之间的注意力权重来生成时序性的区域响应分布，其形式可表示为：
%We assign each body part $g_i$ a learnable part query vector $q_i$, which is used to capture the part's dynamic correlation in the motion sequence. Specifically, the model generates a time-series Regional Response distribution by calculating the attention weight between the regional query and the motion feature, which can be expressed:
Each body part $g_i$ is assigned a learnable part query vector $q_i$, designed to capture the part's dynamic correlations within the motion sequence. Specifically, the model computes attention weights between each part query and the motion features to generate a time-series part response distribution, expressed as:
\begin{equation}
  A_i=Softmax\left(\frac {q_iW_q(F_m'W_k)^T}{\sqrt{\smash[b]{D}}}\right)
\end{equation}
%其中，F_m'\in\R^{T×D} 表示经 BMI 模块融合后的动作特征序列，W_q 与W_k 分别为查询与键的线性投影矩阵，D为特征维度。该操作计算了第 i个身体部位查询向量与所有时间步动作特征之间的加权相似度，并通过 Softmax 函数获得时间维度上的注意力分布 A_i 。
%$F_m'\in\mathbb{R}^{T×D}$ represents the motion feature sequence fused by the BMI module, $W_q$ and $W_k$ are linear projection matrices of queries and keys, respectively, and $D$ is the feature dimension. This operation calculates the weighted similarity between the $i^{th}$ body part query vector and all time-step motion features, and obtains the attention distribution $A_i$ in the time dimension through the Softmax function.
% $F_m' \in \mathbb{R}^{T \times D}$ denotes the motion feature sequence processed by the BMI module, where $W_q$ and $W_k$ are linear projection matrices for queries and keys respectively, and $D$ represents the feature dimension. This operation computes the element-wise similarity between the $i^{th}$ body part query vector and the motion features across all time steps, producing a temporal attention distribution $A_i$ through the Softmax function.
\noindent where $F_m' \in \mathbb{R}^{T \times D}$ is the motion feature sequence, where $W_q$ and $W_k$ are linear projection matrices for queries and keys, respectively. This operation computes the dot-product similarity between the $i^{th}$ body part query vector and the motion features at all time steps, generating a temporal attention distribution $A_i$ via the Softmax function.

%该机制使模型能够显式地区分不同部位的时间动态，进而在后续生成阶段聚焦于语义相关的身体区域。为了进一步捕获不同部位之间的动态协同关系，我们将每个部位的注意力图 Ai作用于语义动作特征 Fm′，得到聚合后的区域特征：
%This mechanism enables the model to explicitly distinguish the temporal dynamics of different parts, and then focus on semantically related body regions in the subsequent generation phase. In order to further capture the dynamic synergy between different parts, we apply the attention map $A_i$ of each part to the semantic motion feature $F_m′$ to obtain the aggregated regional feature:
This mechanism allows the model to explicitly model the temporal dynamics of distinct body parts, facilitating focused attention on semantically relevant parts during the generation stage. To further capture inter-part dependencies, we perform a weighted aggregation of the semantic motion feature $F_m'$ using the attention map $A_i$ for each part, yielding a contextually enriched representation:
\begin{equation}
  z_i=\sum_{\substack{t=1}}^T A_{i,t}F_m'
\end{equation}
%其中A_{i,t}表示第 i 个身体部位在时间步 t上的注意力权重，该过程可视作一个部位特定的加权聚合操作，生成了每个部位在整个时序范围内的语义感知特征表示z_i。通过堆叠所有部位的特征向量，我们得到区域特征序列 Z=[z_1,z_2,z_3,z_4,z_5]，它构成了后续区域建模的输入。
%$A_{i,t}$ denotes the attention weight of the $i^{th}$ body part at time step $t$. This process can be regarded as a part-specific weighted aggregation operation, which generates a semantics-aware feature representation $z_i$ for each part across the entire temporal range. By stacking the feature vectors of all parts, we obtain the regional feature sequence $Z = [z_1, z_2, z_3, z_4, z_5]$, which serves as the input for subsequent regional modeling.
\noindent where $A_{i,t}$ denotes the attention weight of the $i^{th}$ body part at time $t$. This process performs a part-specific weighted aggregation, generating a semantically-informed feature representation $z_i$ for each part over the entire sequence. By concatenating the feature vectors of all five parts, we obtain the part feature sequence $Z = [z_1, z_2, z_3, z_4, z_5]$, which is the input for subsequent part modeling.

%为了捕获不同身体部位之间的动态协同关系，我们在该特征序列上引入轻量级 Transformer 编码器进行空间–时间层面的联合建模：
To capture the dynamic collaborative relationships between different body parts, we introduce a lightweight Transformer encoder on this feature sequence for joint spatial-temporal modeling 
$\hat{Z}=Transformer(Z)$.
%其中每个 Transformer 层由多头自注意力（Multi-Head Self-Attention）与前馈网络（Feed-Forward Network, FFN）组成。自注意力机制允许每个身体部位的表示与其他部位的信息进行交互，从而学习到隐式的运动依赖关系（如手臂动作与躯干姿态的协调调整），同时保持时序连续性与结构一致性。
Each Transformer layer consists of multi-head self-attention and feed-forward network. 

%在获得经过语义协同建模的区域特征\^{Z}后，我们通过一个非线性映射网络将其投影到部位层级的可编辑性空间：
After obtaining the semantic cooperatively modeled part features $\hat{Z}$, we project them into the editability space at the site level through a nonlinear mapping network:
\begin{equation}
  R=\sigma\big(W_2\cdot GELU(W_1\cdot\hat{Z})\big)
\end{equation}
%其中，W_1和W_2为可学习的线性变换矩阵，其中GELU(·) 为高斯误差线性单元，用于增强网络的非线性表达能力并平滑梯度变化。\sigma(·)为Sigmoid 函数，用于约束权重范围。较高的R_{i,t}值意味着该部位在第 t 帧中需要更强的动作调整。结果向量R\in[0,1]^{T×5}表示在每个时间步上各身体部位的可编辑权重。
%$W_1$ and $W_2$ are learnable linear transformation matrices. Among them, $GELU(·)$(Gaussian Error Linear Unit) is used to enhance the network’s nonlinear expression capability and smooth gradient changes. $\sigma(\cdot)$ denotes the Sigmoid function, which serves to constrain the weight range. A higher $R_{i,t}$ value indicates that the corresponding body part requires stronger motion adjustment at the $t^{th}$ frame. The resulting vector $R\in[0,1]^{T\times5}$ represents the editable weights of each body part at every time step.
% $W_1$ and $W_2$ are learnable linear transformation matrices. Here, the GELU (Gaussian Error Linear Unit) activation introduces nonlinearity and promotes smoother gradient flow, while the Sigmoid function $\sigma(\cdot)$ constrains the output to the range $[0,1]$. A higher value of $R_{i,t}$ indicates that the $i^{th}$ body part requires more pronounced motion adjustment at the $t^{th}$ frame. The resulting matrix $R \in [0,1]^{T \times 5}$ thus represents the editable weight of each body part at every time step.
\noindent where $W_1$ and $W_2$ are learnable linear transformation matrices. The GELU activation provides nonlinearity and smooth gradients, while the Sigmoid function $\sigma(\cdot)$ ensures outputs in $[0,1]$. A higher value of $R_{i,t}$ indicates that the $i^{th}$ body part requires stronger motion adjustment at the $t^{th}$ frame, forming the editable weight matrix $R \in [0,1]^{T \times 5}$ for each body part over time.

%最后，我们将预测得到的部位权重作为显式调制因子，作用于动作特征空间，实现语义驱动的区域自适应控制：
%Finally, we take the predicted part weights as explicit modulation factors and apply them to the motion feature space to achieve semantically driven region-adaptive control:
Finally, we leverage the predicted part weights as explicit modulation factors, applying them to the motion feature space through element-wise multiplication to achieve semantically-driven, part-adaptive modulation:
\begin{equation}
  F_m''=F_m'+R\odot MLP(F_m')
\end{equation}
%其中 ⊙ 表示逐元素乘法操作。机制使得 PMM 不仅在训练阶段通过部位相似度曲线获得显式监督，还能在生成阶段实现动态的区域级特征调制，使模型在编辑时能够聚焦于语义相关的身体部位，同时保持非目标区域的动作一致性与自然性。
%where $\odot$ denotes the element-wise multiplication operation. The mechanism enables PMM not only to obtain explicit supervision through the part similarity curve in the training stage, but also to realize dynamic region-level feature modulation in the generation stage, so that the model can focus on semantically related body parts during editing, while maintaining the consistency and naturalness of the motions of non-target areas.
\noindent where $\odot$ denotes element-wise multiplication. This design enables PMM to receive explicit supervision from the part similarity curve during training, while performing dynamic part-level feature modulation during generation. As a result, the model learns to focus on semantically relevant body parts during editing while preserving motion consistency and naturalness in non-target parts.

\subsubsection{Loss functions}
%为了引导模型生成具有物理语义一致性的部位掩码，我们使用前一节构建的部位相似度曲线\bar{S}_{i,t}作为监督信号。PMM 的损失函数由两部分组成：
%To guide the model to generate a site mask with physical semantic consistency, we use the site similarity curve $\bar{S}_{i,t}$ constructed in the previous section as a supervisory signal. PMM's loss function consists of two parts:
To guide the model in generating a part-aware mask with spatial and semantic consistency, we employ the part similarity curve $\bar{S}_{i,t}$ constructed earlier as a supervision signal. The loss function for PMM comprises two components:
\begin{equation}
   \mathcal{L}_{\text{PMM}} = \mathcal{L}_{\text{PSM}} + \lambda_s\mathcal{L}_{\text{smooth}}
\end{equation}

\noindent where the parameter $\lambda_s$ can balance the importance of the smooth term $\mathcal{L}_{\text{smooth}}$, which is set to 0.1.
\begin{equation}
   \mathcal{L}_{\text{PSM}} = \frac 1{NT}\sum_{\substack{i=1}}^N \sum_{\substack{t=1}}^T \lVert 
  R_{i,t}-\bar{S}_{i,t}
    \rVert_2
\end{equation}
%用于约束模型预测的区域权重 R_{i,t}与构建的真实部位相似度曲线\bar{S}_{i,t}在数值分布上保持一致。

%The part weights $R_{i,t}$ that constrain the model predictions are consistent in numerical distribution with the constructed real part similarity curve $\bar{S}_{i,t}$.
The predicted part weights $R_{i,t}$ are optimized to match the distribution of the ground-truth part similarity curve $\bar{S}_{i,t}$ through a regression loss.
%该项损失鼓励模型在时序维度上准确回归各身体部位的编辑强度，从而使所生成的区域掩码能够真实反映目标动作在语义与几何层面的变化趋势。
%This loss encourages the model to accurately regress the editing intensity of each body part in the temporal dimension, enabling the generated part masks to truly reflect the variation trends of the target motion at both the semantic and geometric levels.
This loss ensures precise regression of the temporal editing scope for each body part, enabling the generated part masks to faithfully capture both the semantic intent and kinematic patterns of the target motion, thereby producing part-based targeted and semantically faithful motion edits.

\begin{equation}
\mathcal{L}_{\text{smooth}} = \frac 1{NT}\sum_{\substack{i=1}}^N
\lVert 
  R_{i,1:T-1}-R_{i,2:T}
    \rVert_1
\end{equation}

The smooth term is used to punish the variation of part weights between adjacent frames to ensure the temporal continuity and physical rationality of the prediction results, ensuring the generated motion conforms to the evolution law of human motions.

\subsection{Human motion diffusion model}
%为了将文本指令所描述的编辑意图转化为高质量、平滑的动作序列，我们采用去噪扩散概率模型（DDPM）作为我们方法的核心生成框架。该模型在训练阶段学习从噪声中恢复出干净数据的映射，并在推理阶段通过迭代去噪过程从随机噪声中合成出符合条件分布的运动数据。
To translate the text editing intent into a high-quality and smooth motion sequence, we employ a Denoising Diffusion Probabilistic Model (DDPM)~\cite{ho2020denoising}.
% This model learns to restore clean data from random noise during the training phase and synthesizes motion data that conforms to the conditional distribution through an iterative denoising process during the inference phase.
%在训练阶段，我们以前文得到的调制后动作特征 F_m''作为干净的运动条件。在每一个训练迭代中，我们随机采样一个时间步t，并依据预定义的噪声调度，向F_m''添加相应程度的高斯噪声，得到噪声化后的运动 F_{m}^{(t)}。我们的去噪网络\epsilon_\theta 以 F_{m}^{(t)}、时间步t以及来自BMI模块的增强文本特征F_{t'}作为条件输入，其目标是预测所添加的噪声。我们采用均方误差（L2）损失来监督这一预测过程，其目标函数定义为：
%During the training phase, we use the modulated motion feature $F_m''$ as the clean motion condition. In each training iteration, we randomly sample a timestep $t$, according to a predefined noise schedule, add a corresponding level of Gaussian noise to $F_m′′$ to obtain the noised motion $F_{m}^{(t)}$. Our denoising network $\epsilon_\theta$ takes $F_{m}^{(t)}$, the time step $t$, and the enhanced text feature $F_{t}'$ from the BMI module as conditional inputs. Its objective is to predict the added noise. We use the mean squared error (L2) loss to supervise this prediction process, with the objective function defined as:
During training, we condition the diffusion process on the modulated motion feature $F_m''$. At each training iteration, we sample a random timestep $t$ from a predefined noise schedule and corrupt $F_m''$ with a corresponding level of Gaussian noise, obtaining the noisy motion representation $F_m^{(t)}$. The denoising network $\epsilon_\theta$ is then conditioned on $F_m^{(t)}$, the timestep $t$, and the enhanced text feature $F_t'$ from BMI, and is trained to predict the added noise. The training objective is formulated as a mean squared error (L2) loss:
\begin{equation}
   \mathcal{L}_{\text{}} =\mathbb{E}_{t,F_{m}'',\epsilon\thicksim\mathcal{N}(0,I)}
\lVert 
  \epsilon-\epsilon_\theta(F_m^{(t)},t,F_t')
    \rVert_2
\end{equation}
%通过这一优化目标，模型能够逐步学习在文本条件F_{t'}的引导下，从任意噪声状态中重建出符合语义的、高质量的动作特征。
Through this optimization objective, the model gradually learns to reconstruct semantically accurate and high-quality motion features from any noisy state, guided by the text condition $F_{t}'$.
%在推理编辑阶段，我们从一个标准高斯分布中采样得到初始噪声运动 F_{m}^{(N)}。随后，我们执行N次迭代去噪：在每一步中，去噪网络\epsilon_\theta 根据当前噪声运动F_{m}^{(n)}、当前时间步n 和文本条件 F_{t'}预测出噪声估计，并依据DDPM的采样规则计算出更干净的运动 F_{m}^{(n-1)}。经过这一系列精炼步骤后，最终的输出F_{m}^{(0)} 即为由文本指令P所驱动的、编辑后的目标动作序列 M_{tgt}。
% During the inference and editing stage, we first sample an initial noisy motion $F_{m}^{(N)}$ from a standard Gaussian distribution. We then perform $N$ iterative denoising steps: at each step, the denoising network $\epsilon_\theta$ predicts a noise estimate based on the current noisy motion $F_{m}^{(n)}$, the current time step $n$, and the text condition $F_{t}'$. Following the sampling rule of DDPM, a cleaner motion $F_{m}^{(n-1)}$ is computed. After this series of refinement steps, the final output $F_{m}^{(0)}$ represents the edited target motion sequence $M_{tgt}$, driven by the text instructions.
During inference, the model iteratively denoises an initial Gaussian noise sample $F_{m}^{(N)}$ over N steps, conditioned on the text feature $F_{t}'$, to generate the final motion sequence.
\section{Results} 
\label{sec:results}

\subsection{Experiment setup} 
\noindent \textbf{The dataset.}
%MotionFix 数据集是首个面向语言驱动 3D 动作编辑任务的大规模数据集。它由“源动作—目标动作—文本指令”三元组组成，用于学习如何根据语言描述对给定动作进行局部或整体修改。为保证语义一致性，作者利用 TMR 运动嵌入空间从 AMASS 动作捕捉库中检索语义相近的动作对，并通过人工标注为其撰写简洁的差异描述。最终，MotionFix 共包含 6,730 条高质量三元组样本，涵盖不同类型的编辑，如身体部位修改、动作速度调整与风格变换。数据集按 8∶0.5∶1.5 比例划分为训练、验证和测试集。
MotionFix is the first large-scale dataset dedicated to text-driven 3D motion editing tasks~\cite{athanasiou2024motionfix}. 
% To ensure semantic consistency, the authors use the TMR motion embedding space~\cite{petrovich2023tmr} to retrieve semantically similar motion pairs from the AMASS motion capture library~\cite{mahmood2019amass} and compose concise difference descriptions for them through manual annotation.
MotionFix contains 6,730 high-quality source-text-motion triple samples. It covers various editing types, such as body part modification, motion speed adjustment, and style transformation. We divided MotionFix into training, validation, and test sets randomly with a proportion of 8:0.5:1.5.

\begin{table*}[h]
\centering
\caption{Comparison with SOTA text-based motion editing methods on the MotionFix~\cite{athanasiou2024motionfix} benchmark. We calculate the standard indicator according to MotionFix. Bold represents the best, underline represents the second best, ↑ / ↓ represents higher / lower values are better.}
\begin{tabular}{c cccc cccc}
\hline
\multirow{2}{*}{\textbf{Methods}} & \multicolumn{4}{c}{\textbf{Generated-to-Target (Batch)}} & \multicolumn{4}{c}{\textbf{Generated-to-Target (Test Set)}} \\
\cline{2-9}
 & \textbf{R@1$\uparrow$} & \textbf{R@2$\uparrow$} & \textbf{R@3$\uparrow$} & \textbf{AvgR$\downarrow$}\,\,\,\,\,\, & \textbf{R@1$\uparrow$} & \textbf{R@2$\uparrow$} & \textbf{R@3$\uparrow$} & \textbf{AvgR$\downarrow$} \\
\hline
GT      & 100.0 & 100.0 & 100.0 & 1.00  & 64.36 & 88.75 & 95.56 & 1.74  \\
MDM\cite{tevet2023human}    & 4.03  & 7.56  & 10.48 & 15.55 & 0.10  & 0.10  & 0.10  & -     \\
MDM-BP\cite{athanasiou2024motionfix}  & 39.10 & 50.09 & 54.84 & 6.46  & 8.69  & 14.71 & 18.36 & 180.99\\
TMED\cite{athanasiou2024motionfix}    & 62.90 & 76.51 & 83.06 & 2.71  & 14.51 & 21.72 & 28.73 & 56.63 \\
MotionRefit\cite{jiang2025dynamic} & 66.63 & 80.05 & 84.98 & 2.64 &   14.13  &   23.52   &   30.53   &   54.06  \\
SimMotionEdit\cite{li2025simmotionedit} & \underline{70.62} & \underline{82.92} & \underline{88.12} & \underline{2.38} & \underline{25.49} & \underline{39.33} & \underline{49.21} & \underline{23.49} \\
\mname\ (ours)\,\,\,\,    & \textbf{73.96} & \textbf{85.83} & \textbf{90.21} & \textbf{1.92}  & \textbf{27.27} & \textbf{45.06} & \textbf{53.36} & \textbf{16.24} \\
\hline
\end{tabular}
\label{tab:SOTAcompare}
\end{table*}

\noindent \textbf{Evaluation metrics.}
%为了客观评估生成动作的编辑质量，我们沿用了SimMotionEdit的评估方法。具体而言，利用预训练的 TMR模型作为特征提取器，将每段动作编码到嵌入空间中。然后计算生成动作与目标动作（generated-to-target retrieval, G→T）或与源动作（generated-to-source retrieval, G→S）之间的特征相似度，并在测试集中执行检索操作。我们报告标准召回指标 R@1、R@2、R@3 以及平均排名 AvgR，其中 R@k 表示目标动作出现在前 k 个检索结果中的比例，AvgR 则衡量整体排名表现。同时，我们在消融研究阶段使用MotionCritic评分（M-scores）来评估动作的保真度，我们沿用SimMotionEdit的方法让生成的动作去匹配作者预训练的模型输入，以实现实验结果的对比。
% In order to objectively evaluate the editing quality of the generated motion, we follow the evaluation method of SimMotionEdit \cite{li2025simmotionedit}. Specifically, the pre-trained TMR model is used as a feature extractor to encode each motion into the embedding space. Then calculate the feature similarity between the generated motion and the target motion (generated-to-target return, g→t) or the source motion (generated-to-source return, g→s), and perform the retrieval operation in the test set. We report the standard recall indicators R@1, R@2, R@3 and the average ranking AvgR, where R@k represents the proportion of the target motion that appears in the top k search results, and AvgR measures the overall ranking performance.At the same time, we use MotionCritic score (M-scores) to evaluate the fidelity of actions in the ablation research phase \cite{wang2024aligning}. We continue to use the SimMotionEdit method to make the generated actions match the author's pre-trained model input to achieve the comparison of experimental results.
To fairly evaluate the quality of the generated motion, we follow the evaluation protocol of SimMotionEdit~\cite{li2025simmotionedit}. More precisely, a pre-trained TMR model~\cite{petrovich2023tmr} is first employed as a feature extractor to encode each motion into an embedding space. Then, We compute the feature similarity between the generated motion and the target motion (generated-to-target, g→t) or the source motion (generated-to-source, g→s), and perform retrieval within the test set. Standard recall metrics (R@1, R@2, R@3) and the average rank (AvgR) are recorded. R@k denotes the proportion of target motions retrieved among the top‑k results, while AvgR reflects the overall ranking performance. Additionally, in the ablation study phase, we adopt the M‑Score~\cite{wang2024aligning} to evaluate motion fidelity. We follow the approach used in SimMotionEdit to make the generated motions compatible with the pre-trained model input provided by the authors.

\noindent \textbf{Implementation details.}
%我们基于 MotionFix 数据集进行训练，采用 SMPL-H 模型进行人体姿态参数化。文本特征由预训练的 CLIP ViT-L/14 模型提取，并在本任务中冻结权重。动作特征经由motion Encoder 编码为 512 维潜空间表示，经由所提出的 BMI 模块 实现语言与动作的双向语义对齐，再由 PMM模块 预测时序部位编辑权重，并输入至 扩散生成网络 进行动作重建与编辑。
% We train based on the MotionFix \cite{athanasiou2024motionfix} dataset and use the SMPL-H model for human pose parameterization. Text features are extracted by a pre-trained CLIP ViT-L/14 model and the weights are frozen in this task \cite{radford2021learning}. motion features are encoded into a 512-dimensional submersible space representation via a Motion Encoder, Bidirectional Semantic alignment between language and motion is achieved through the proposed BMI module, and then the PMM module predicts the editing weights of time series parts and inputs them into the diffusion generation Network for motion reconstruction and editing.
We train our model using the MotionFix \cite{athanasiou2024motionfix} dataset and represent human poses with the SMPL-H model. Text features are extracted by a pre-trained CLIP \cite{radford2021learning} ViT-L/14 model, whose weights remain frozen throughout the task. Motion features are encoded into a 512-dimensional embedding space via a motion encoder. Bidirectional semantic alignment between language and motion is achieved using the proposed BMI module. Subsequently, the PMM module predicts temporal editing weights, which are fed into the diffusion-based generation network for motion reconstruction and editing.

%PMM模块采用两层 Transformer 编码结构，隐藏维度为 256，注意力头数为 4。扩散模型遵循 DDPM 框架，设定训练步数为 300，采样步数为 200。模型在单张 NVIDIA RTX 3090 GPU 上以 batch size 128 训练 1500 个 epoch，优化器为 AdamW，初始学习率为 1×10⁻⁴，并采用线性 warm-up 策略。
% PMM module adopts two-layer Transformer coding structure,The hidden dimension is 256 and the number of attention heads is 4. The diffusion model follows the DDPM framework \cite{ho2020denoising}, with 300 training steps and 200 sampling steps. The model is trained on a single NVIDIA RTX 3090 GPU with batch size 128 for 1500 epochs,The Optimizer is AdamW \cite{loshchilov2017decoupled}, with an initial learning rate of 1 × 10.
The PMM module employs a two-layer Transformer encoder structure with 256 hidden dimensions and 4 attention heads. The diffusion model is built on the DDPM~\cite{ho2020denoising}, using 300 training steps and 200 sampling steps. Training is conducted on a single NVIDIA RTX 3090 GPU with a batch size of 128 and 1500 epochs. The model is optimized using AdamW~\cite{Loshchilov2017DecoupledWD}, with an initial learning rate $10^{-4}$.

\subsection{Comparison with SOTA methods} 
%我们在表1中报告了在MotionFix数据集上的主要实验结果。为保障统计可靠性，所有指标均基于20次重复实验计算其平均值与95%置信区间。由于本方法基于SimMotionEdit进行改进，因此我们重点对比了与该方法的性能差异。实验结果表明，无论是在Batch size为32的采样设置下，还是在完整测试集上，本方法在各项评价指标中均取得了最佳性能。
% We report the main experimental results on the MotionFix dataset \cite{athanasiou2024motionfix}in Table \ref{tab:SOTAcompare}. In order to ensure statistical reliability, all indicators were calculated based on 20 repeated experiments and their average and 95\% confidence interval. Since this method is improved based on SimMotionEdit \cite{li2025simmotionedit}, we focus on comparing the performance difference with this method.The experimental results show that the method achieves the best performance in all evaluation indicators, whether in the sampling setting of Batch size 32 or on the complete test set.
Table~\ref{tab:SOTAcompare} shows our evaluation on the MotionFix~\cite{athanasiou2024motionfix} benchmark. To ensure statistical reliability, all metrics are computed over 20 repeated runs and reported as the mean with a 95\% confidence interval. As our method is built upon SimMotionEdit~\cite{li2025simmotionedit}, we focus primarily on comparing performance differences against this baseline. Experimental results demonstrate that our approach achieves the best performance across all evaluation metrics, both under the batch size 32 sampling setting and on the full test set.

\begin{table*}[h]
\centering
\caption{Qualitative Results. We compare our method with SimMotionEdit \cite{li2025simmotionedit}. We outline the limitations of SimMotionEdit using rectangular boxes and annotate the corresponding colors in the prompt. }
\label{tab:comp_visual}
\begin{tabular}{c@{\hspace{0.005\linewidth}}c@{\hspace{0.005\linewidth}}c@{\hspace{0.005\linewidth}}c}
\hline
\textbf{Source} & \textbf{Ground truth} & \textbf{SimMotionEdit} & \textbf{Ours} \\
\hline
\multicolumn{4}{l}{Prompt: {\color{blue}\textbf{lower arms slightly faster}}. perform the final motion{\color{green}\textbf{ only with the right arm}}, tapping in front, to your left and right} \\
\includegraphics[width=0.245\linewidth]{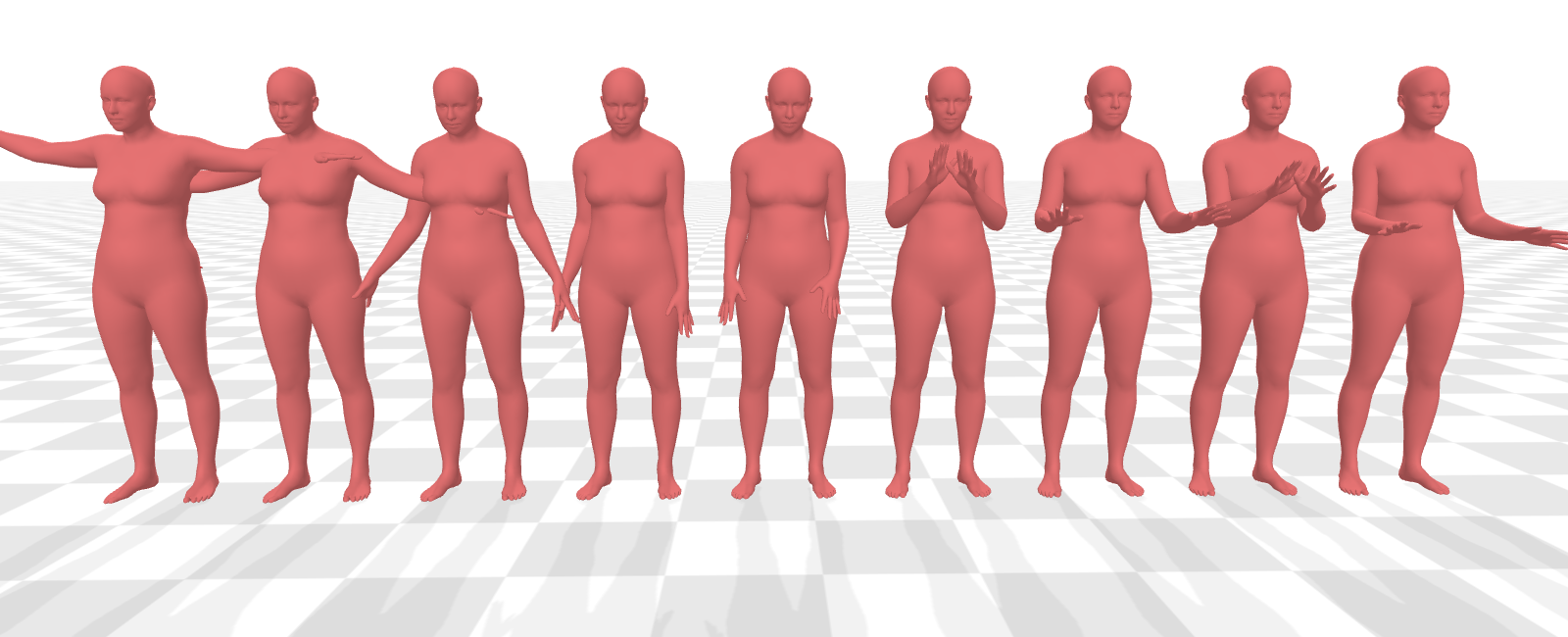} & \includegraphics[width=0.245\linewidth]{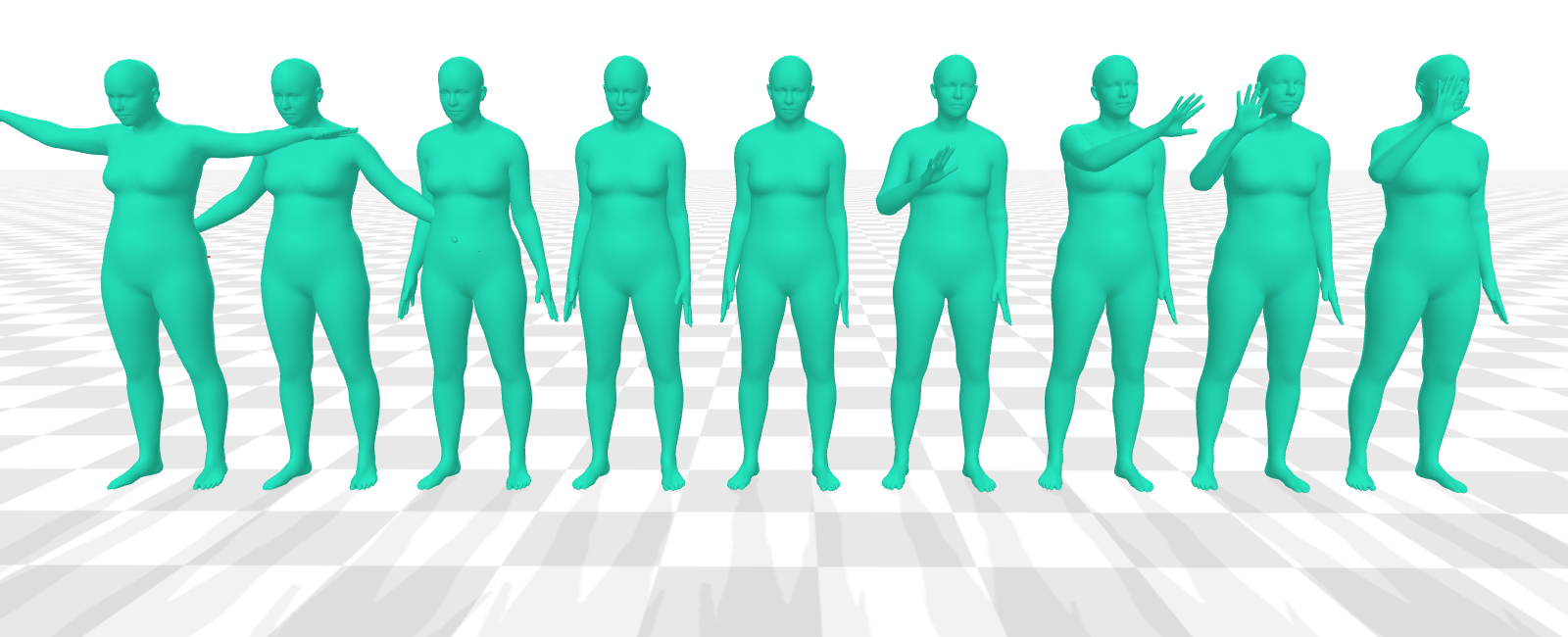} & \includegraphics[width=0.245\linewidth]{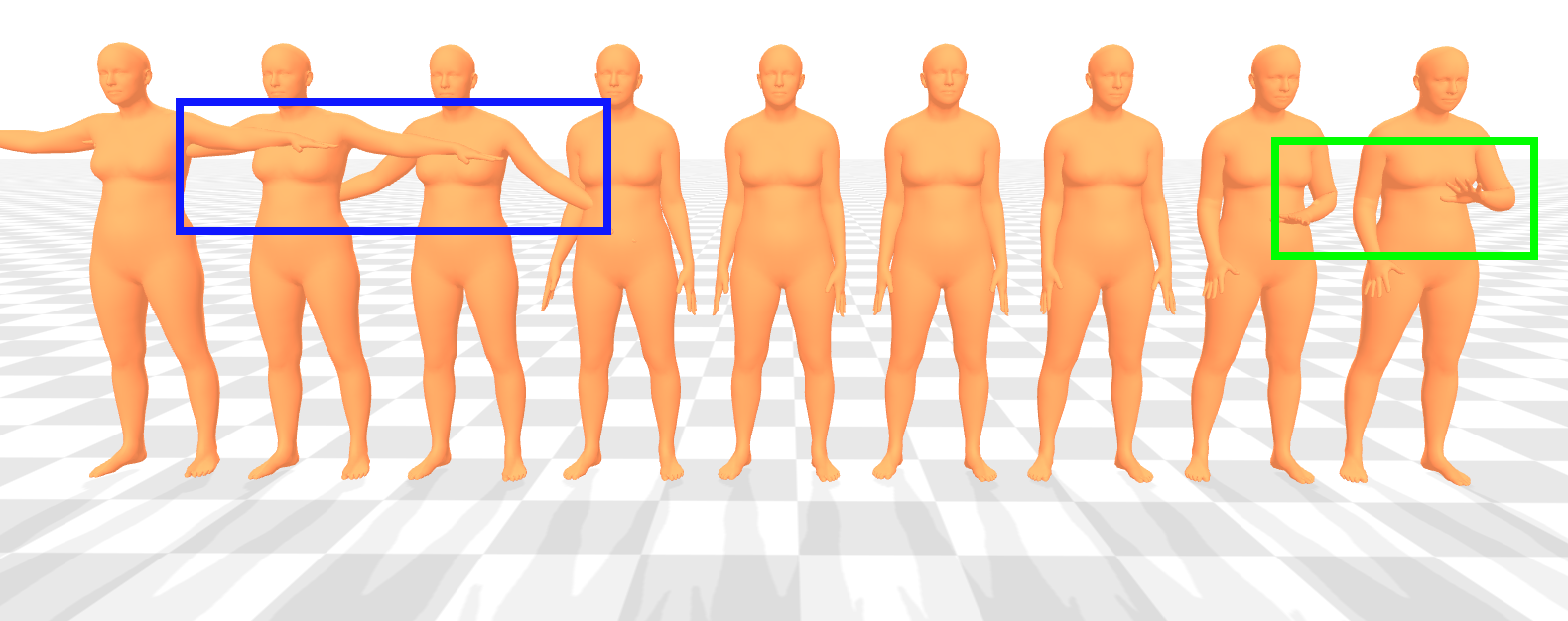} & \includegraphics[width=0.245\linewidth]{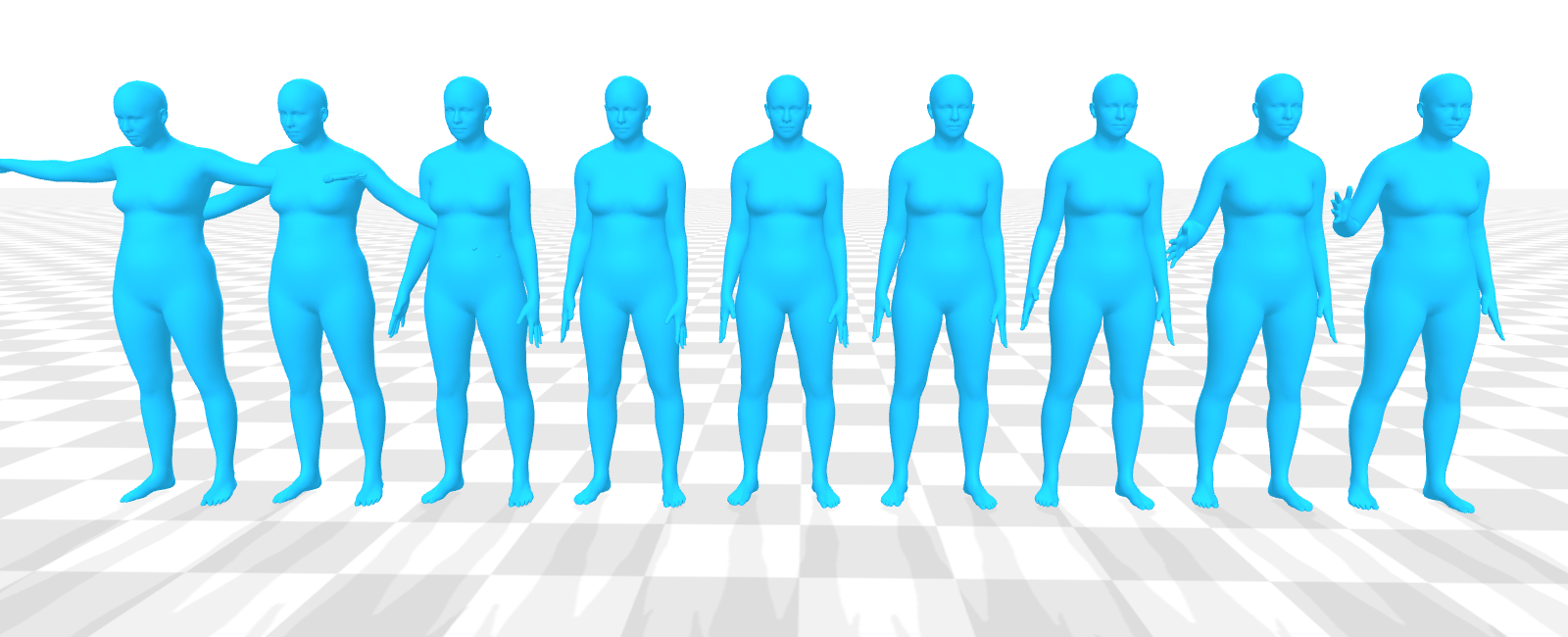} \\
\hline
\multicolumn{4}{l}{Prompt: {\color{green}\textbf{lower arms}} {\color{blue}\textbf{slower and later}}} \\
\includegraphics[width=0.245\linewidth]{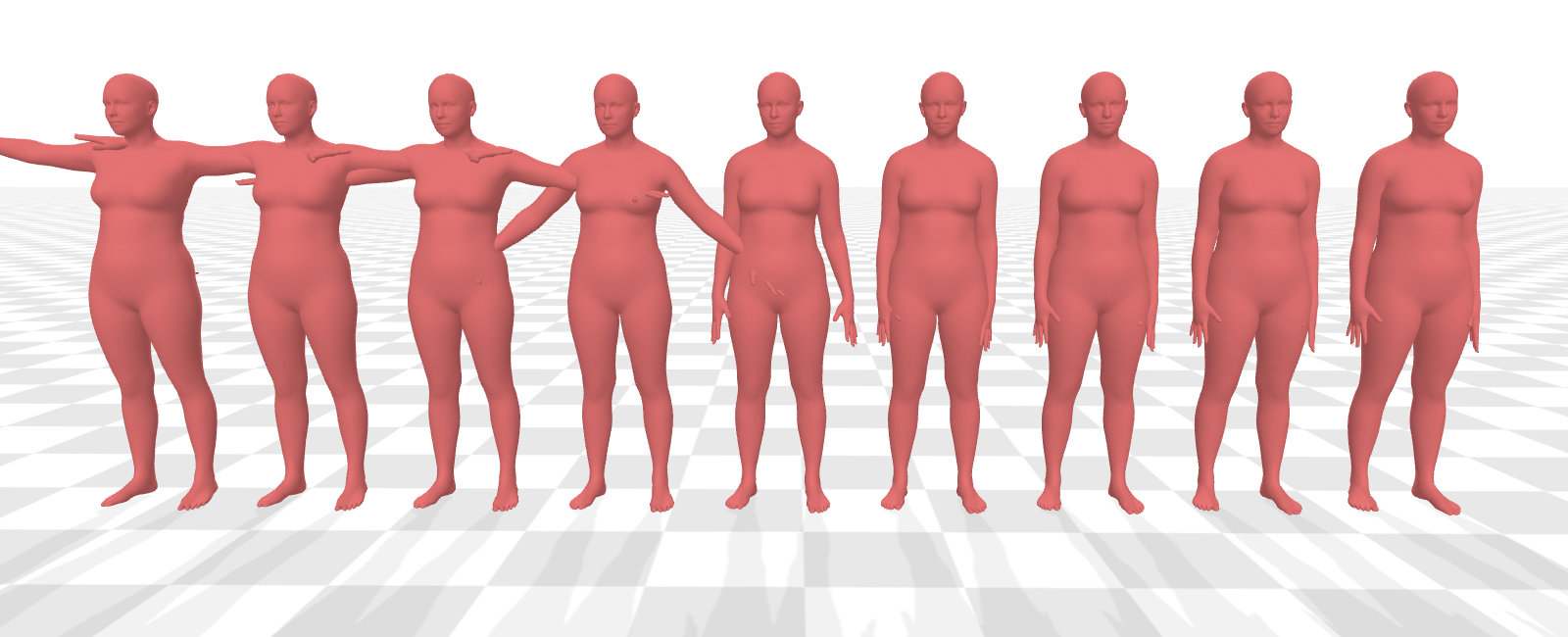} & \includegraphics[width=0.245\linewidth]{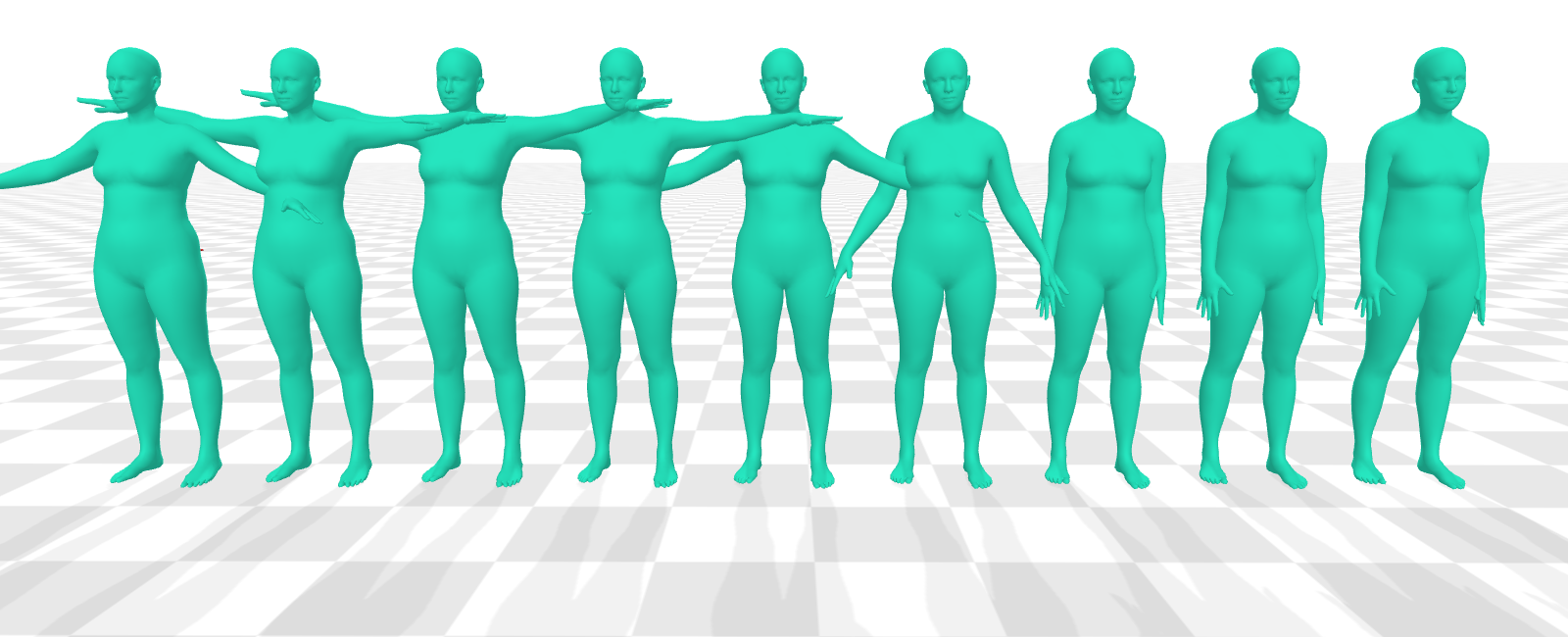} & \includegraphics[width=0.245\linewidth]{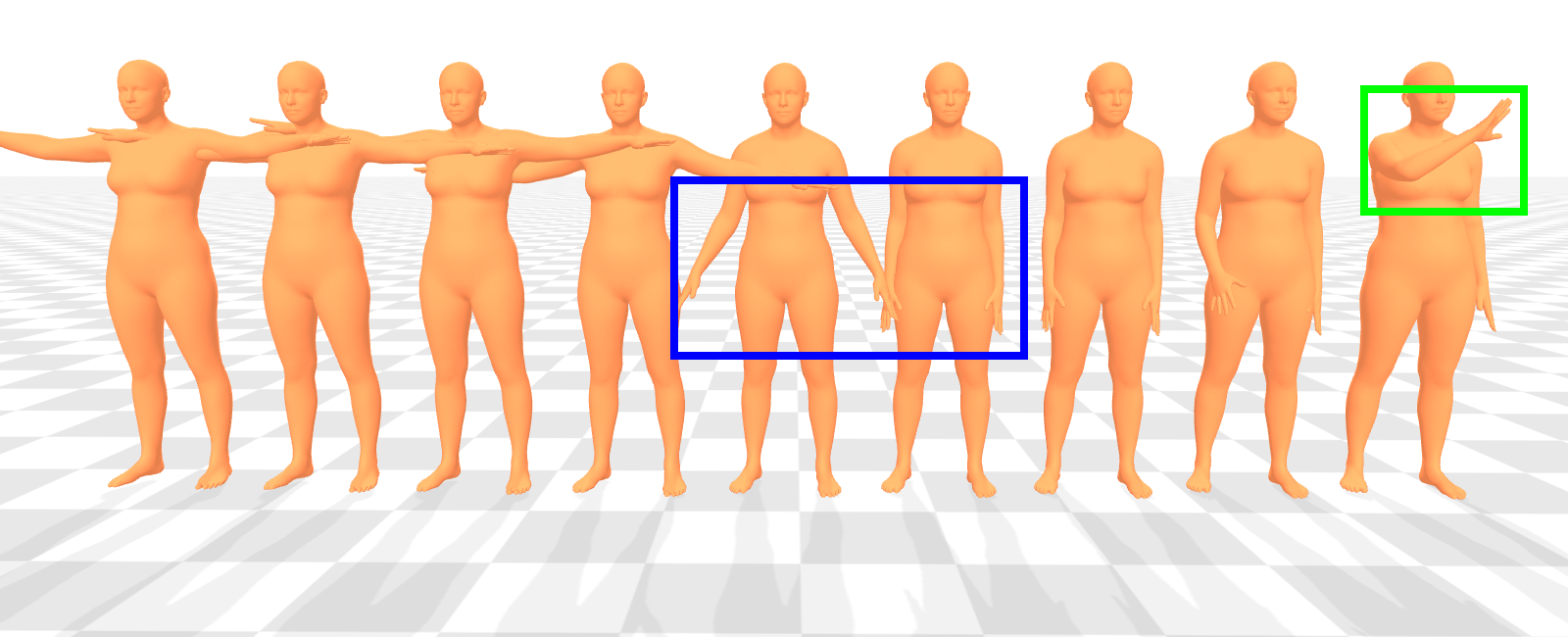} & \includegraphics[width=0.245\linewidth]{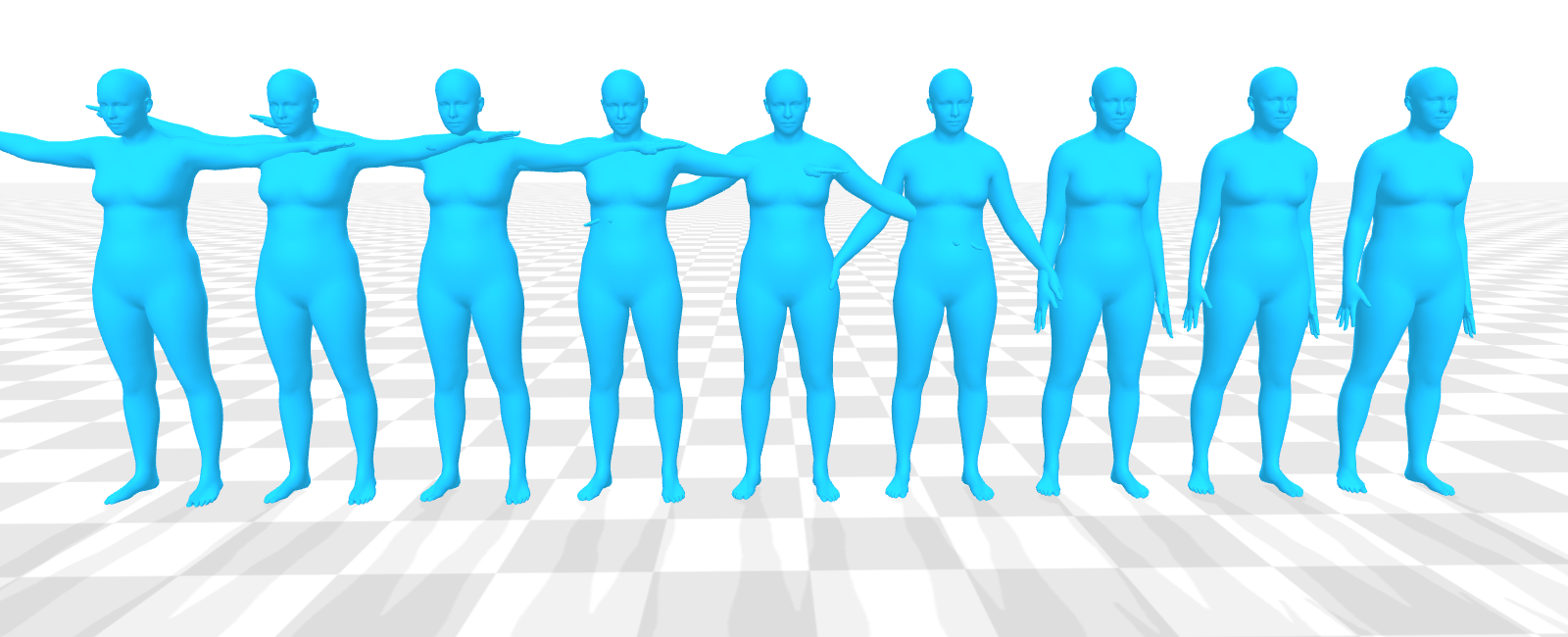} \\
\hline
\multicolumn{4}{l}{Prompt: do the {\color{blue}\textbf{same}} moves in the opposite side } \\
\includegraphics[width=0.245\linewidth]{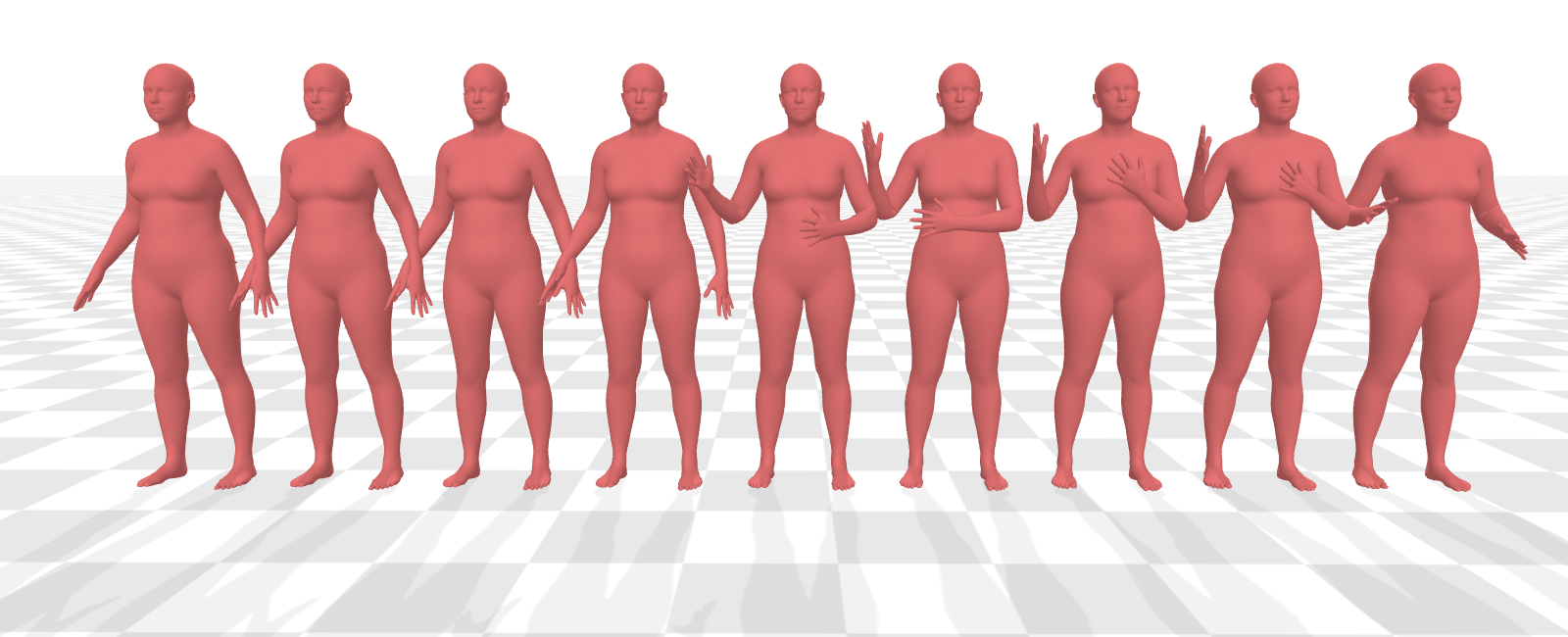} & \includegraphics[width=0.245\linewidth]{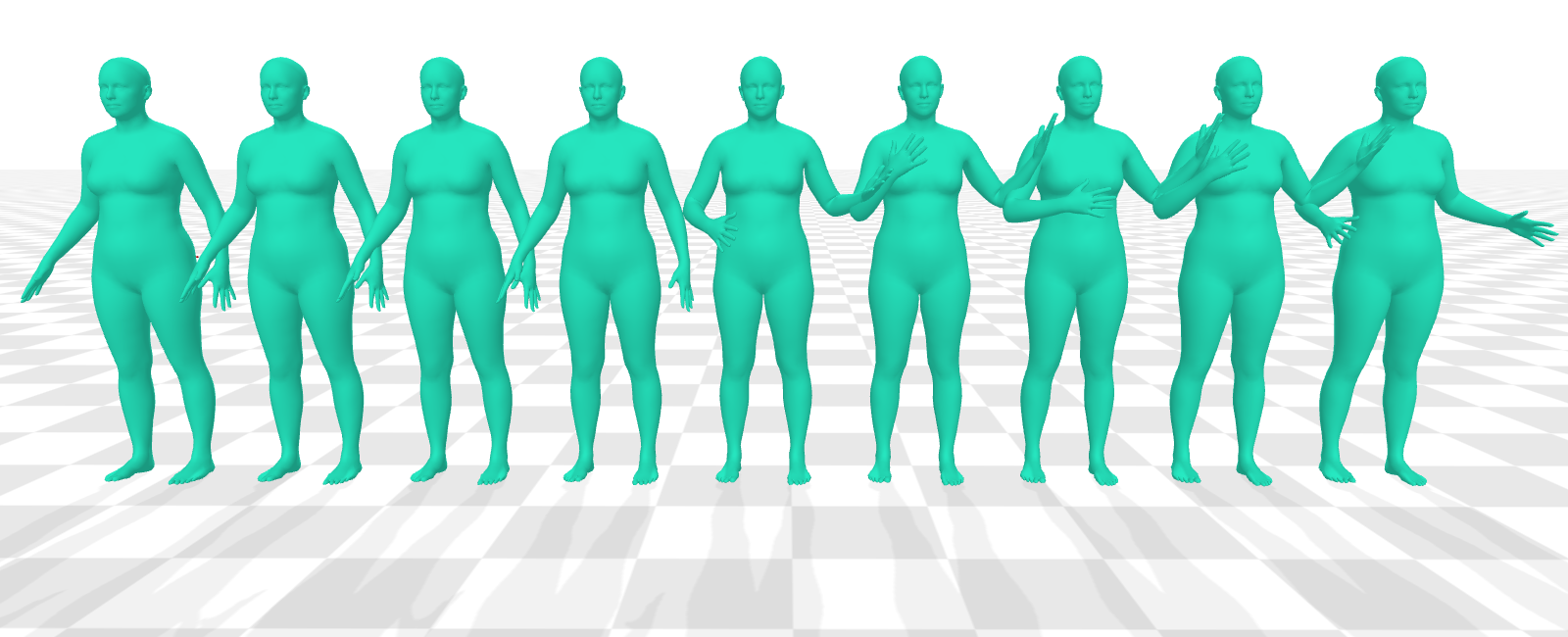} & \includegraphics[width=0.245\linewidth]{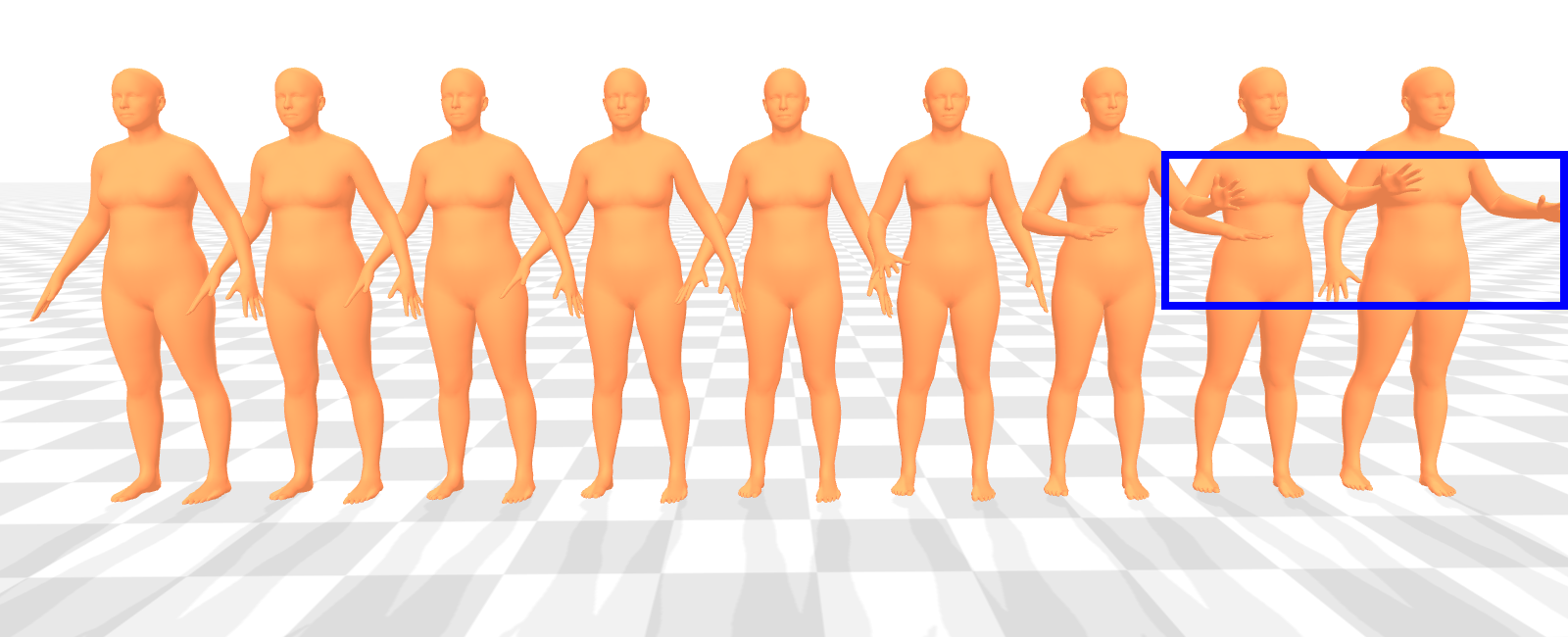} & \includegraphics[width=0.245\linewidth]{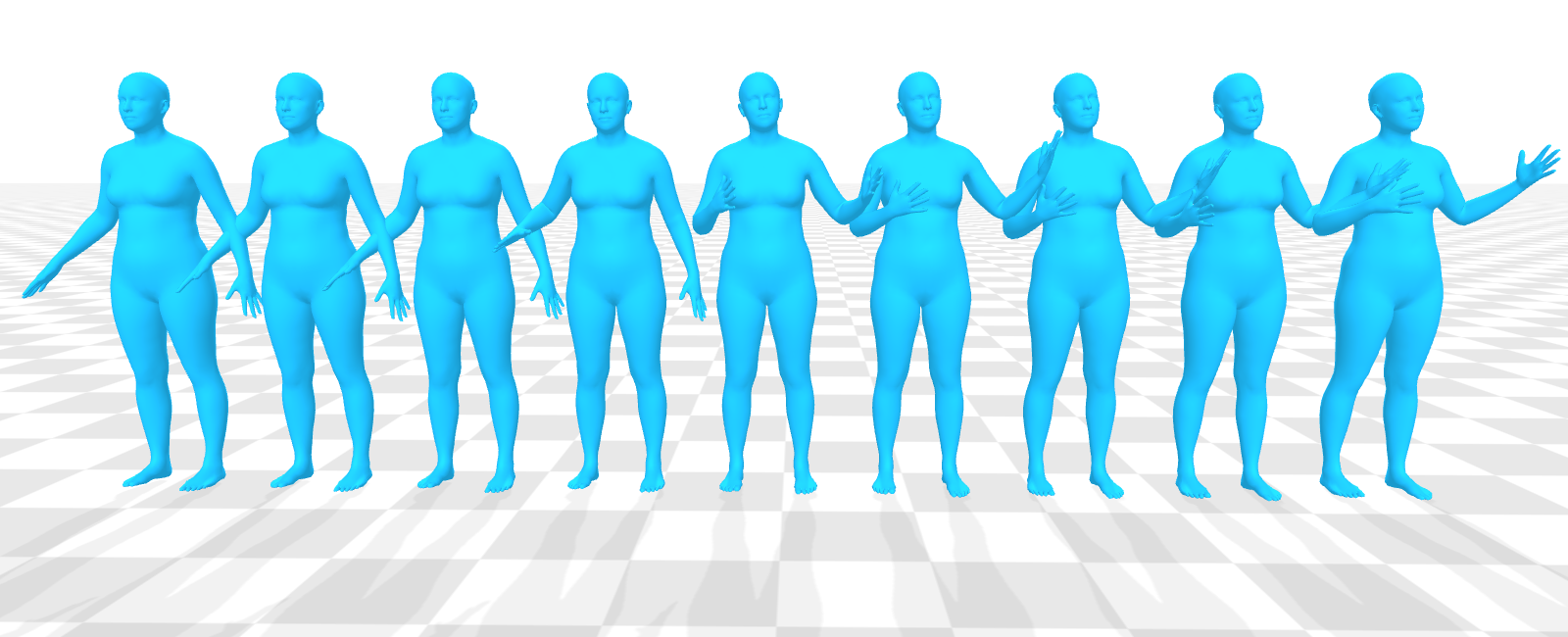} \\
\hline
\multicolumn{4}{l}{Prompt: {\color{blue}\textbf{not}} try to {\color{blue}\textbf{stand up}} keep going } \\
\includegraphics[width=0.245\linewidth]{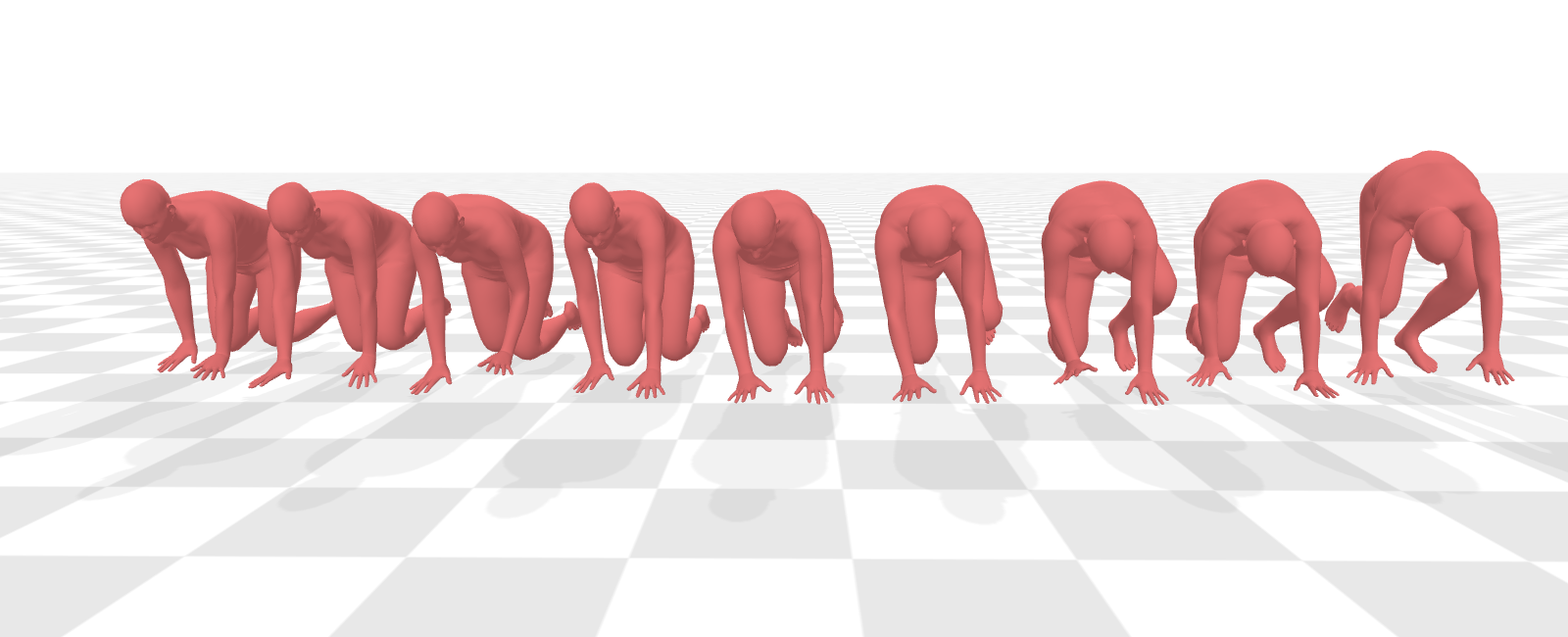} & \includegraphics[width=0.245\linewidth]{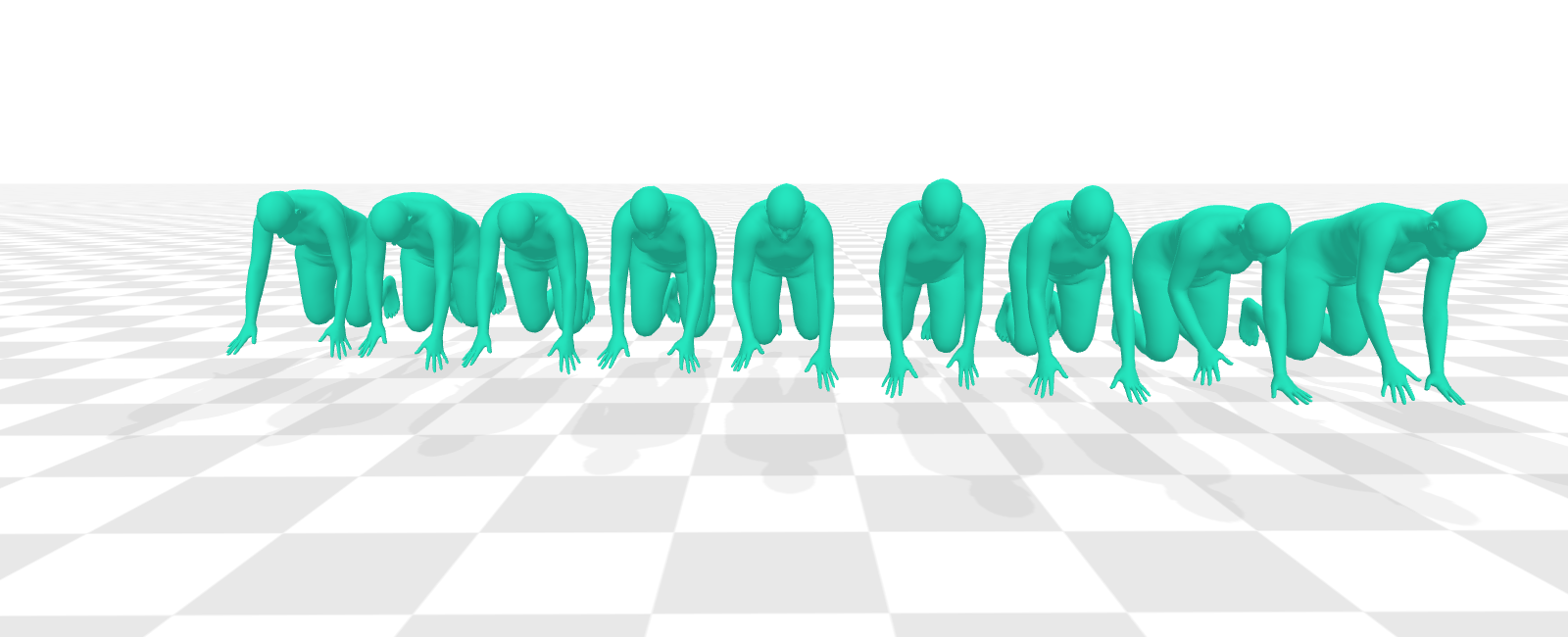} & \includegraphics[width=0.245\linewidth]{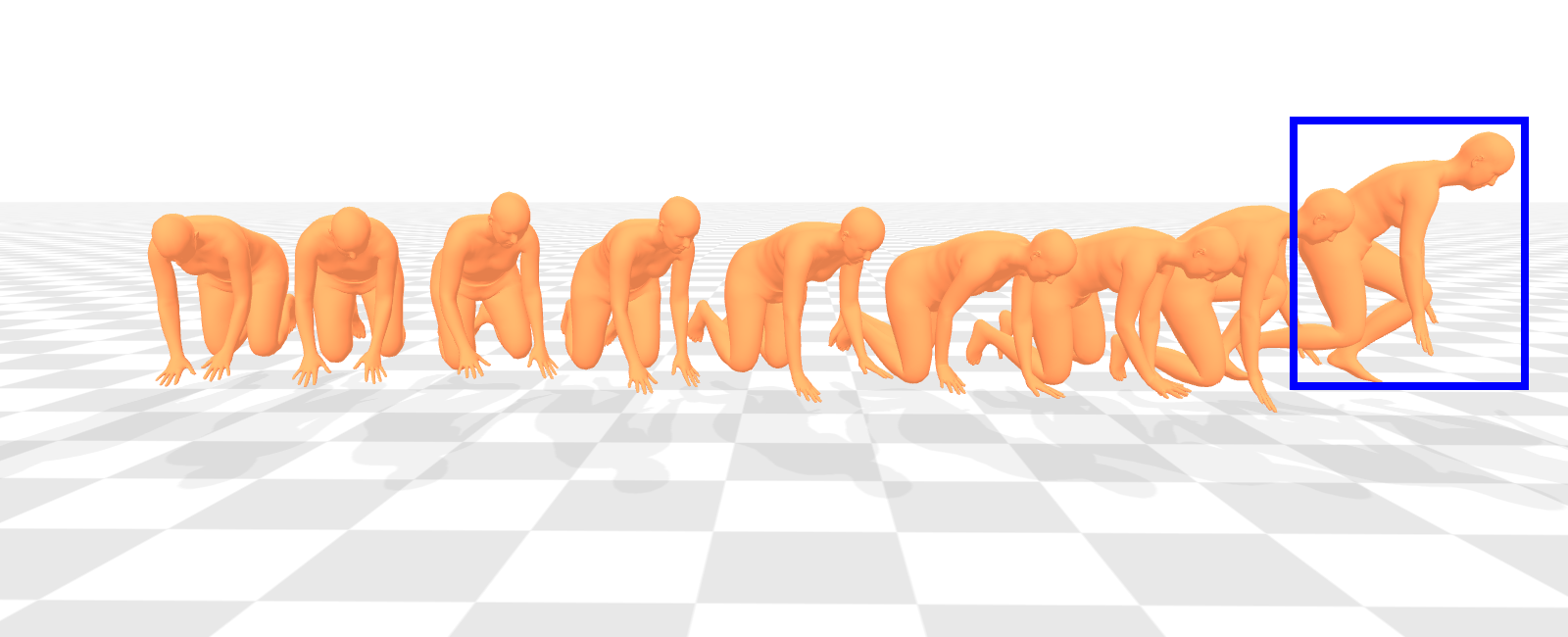} & \includegraphics[width=0.245\linewidth]{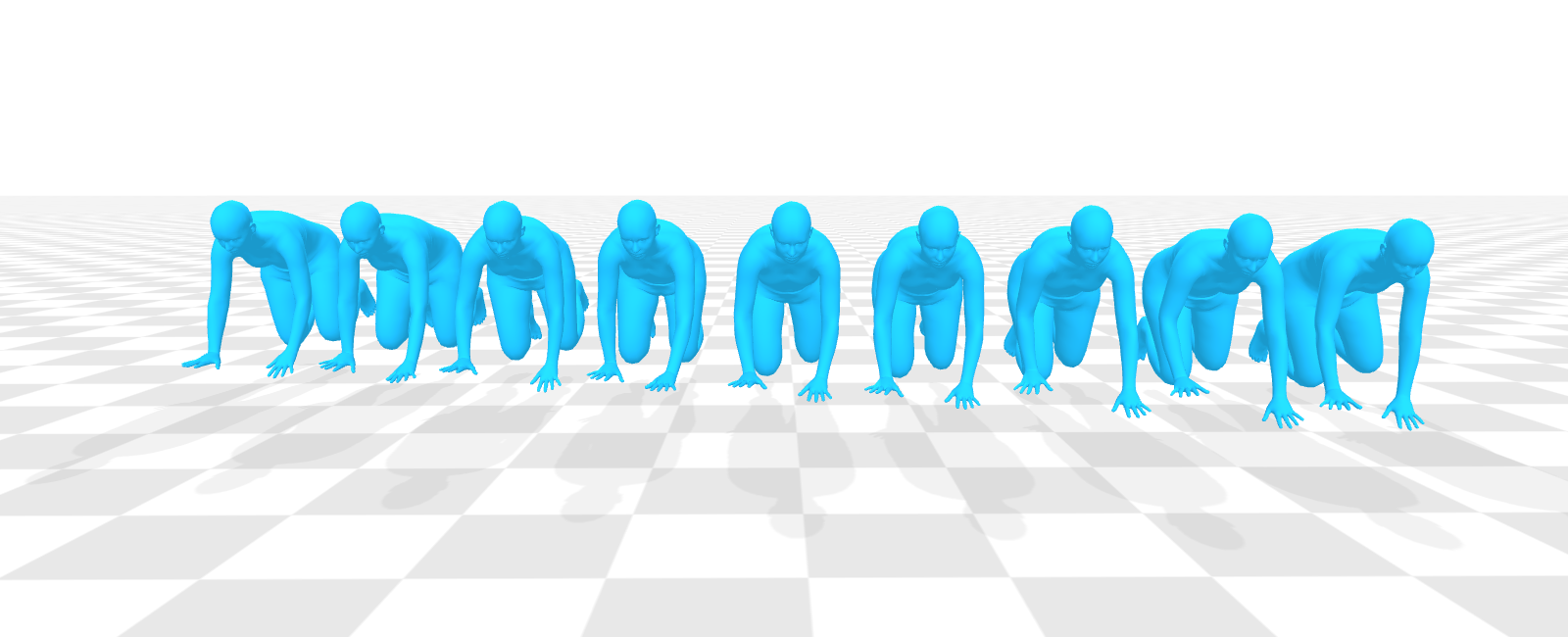} \\
\hline
\multicolumn{4}{l}{Prompt: cross legs faster and {\color{blue}\textbf{get up after a bit}} } \\
\includegraphics[width=0.245\linewidth]{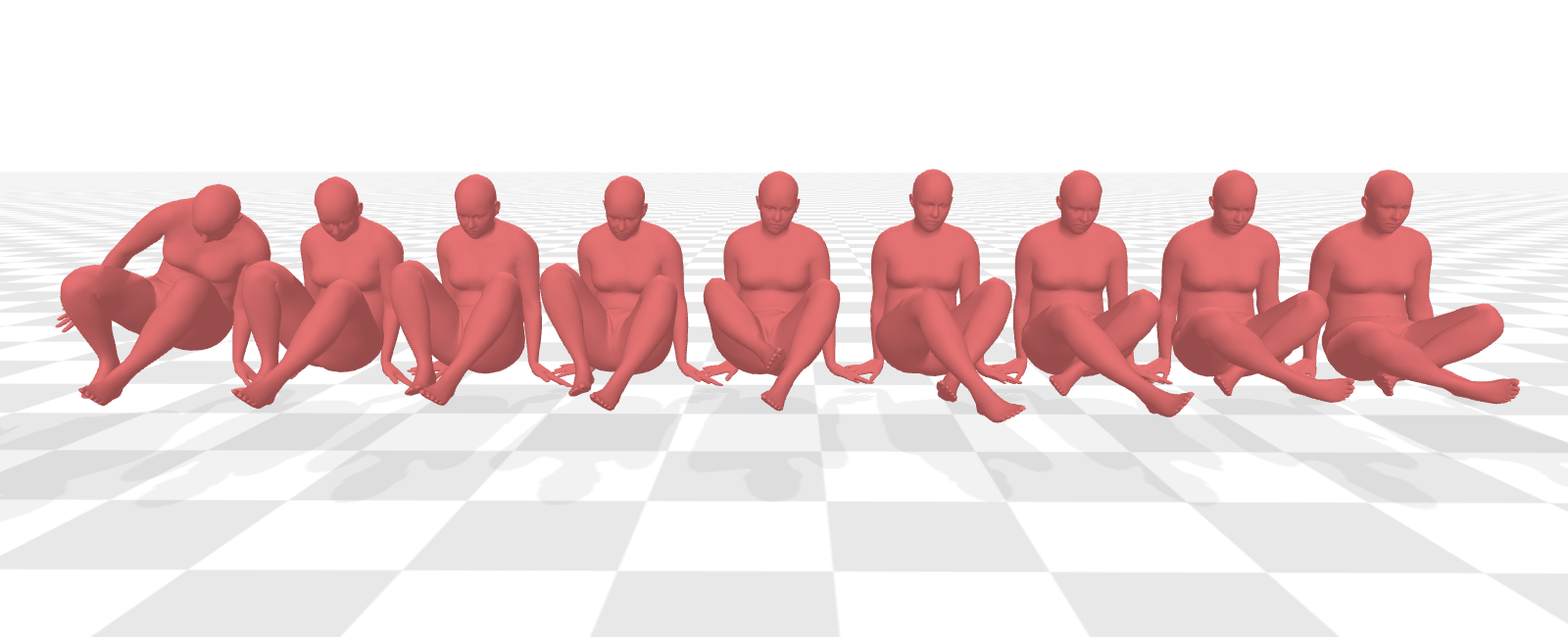} & \includegraphics[width=0.245\linewidth]{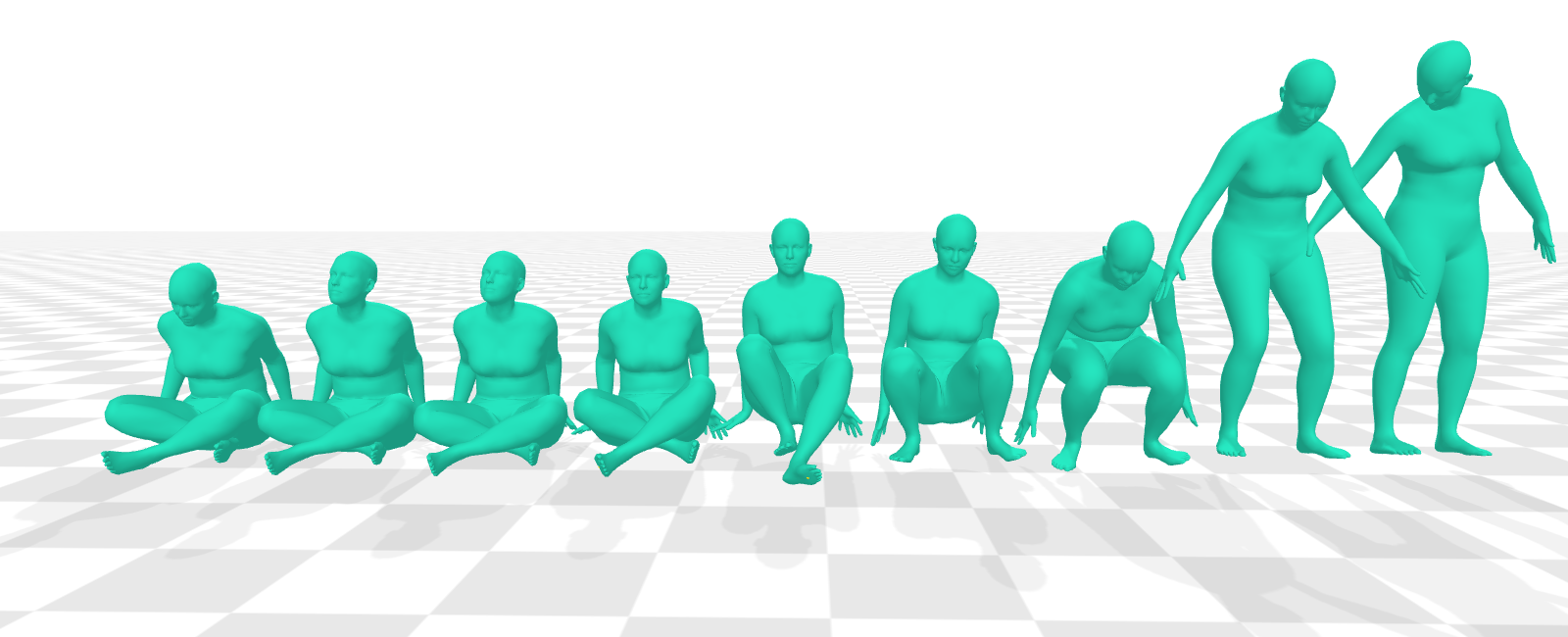} & \includegraphics[width=0.245\linewidth]{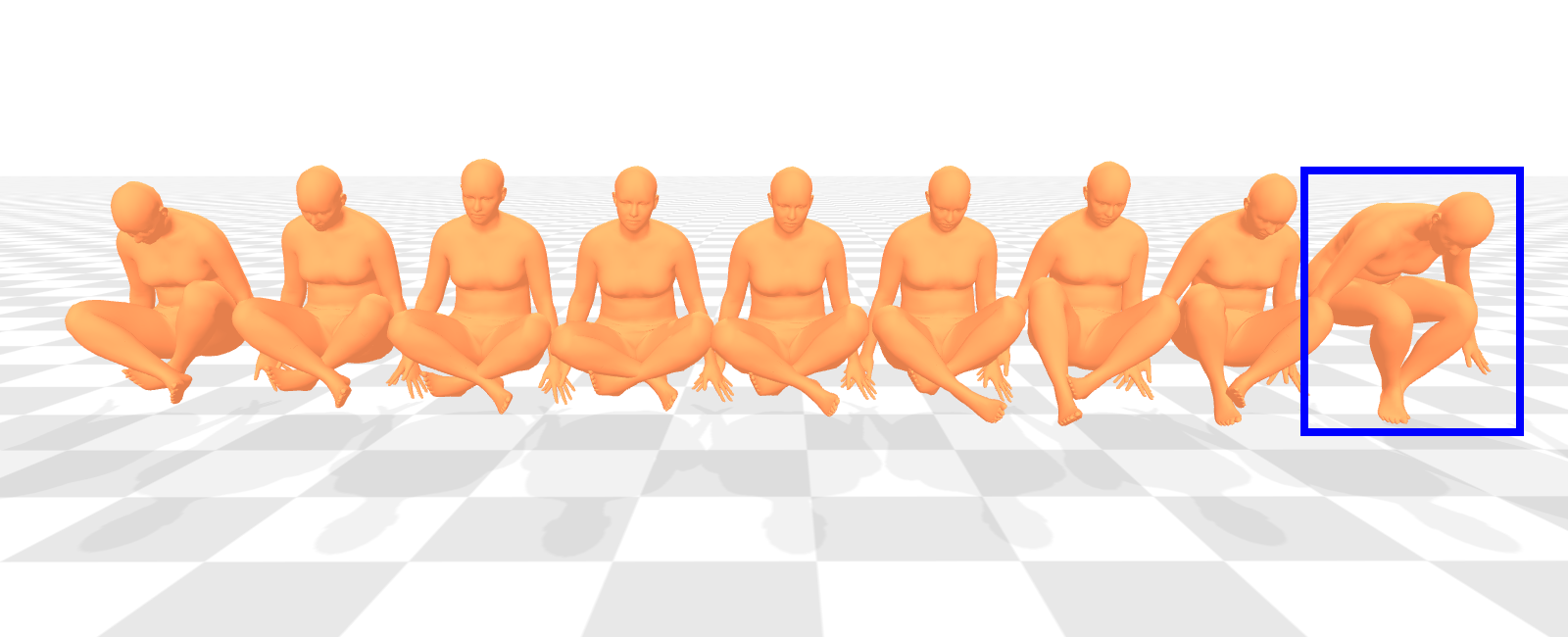} & \includegraphics[width=0.245\linewidth]{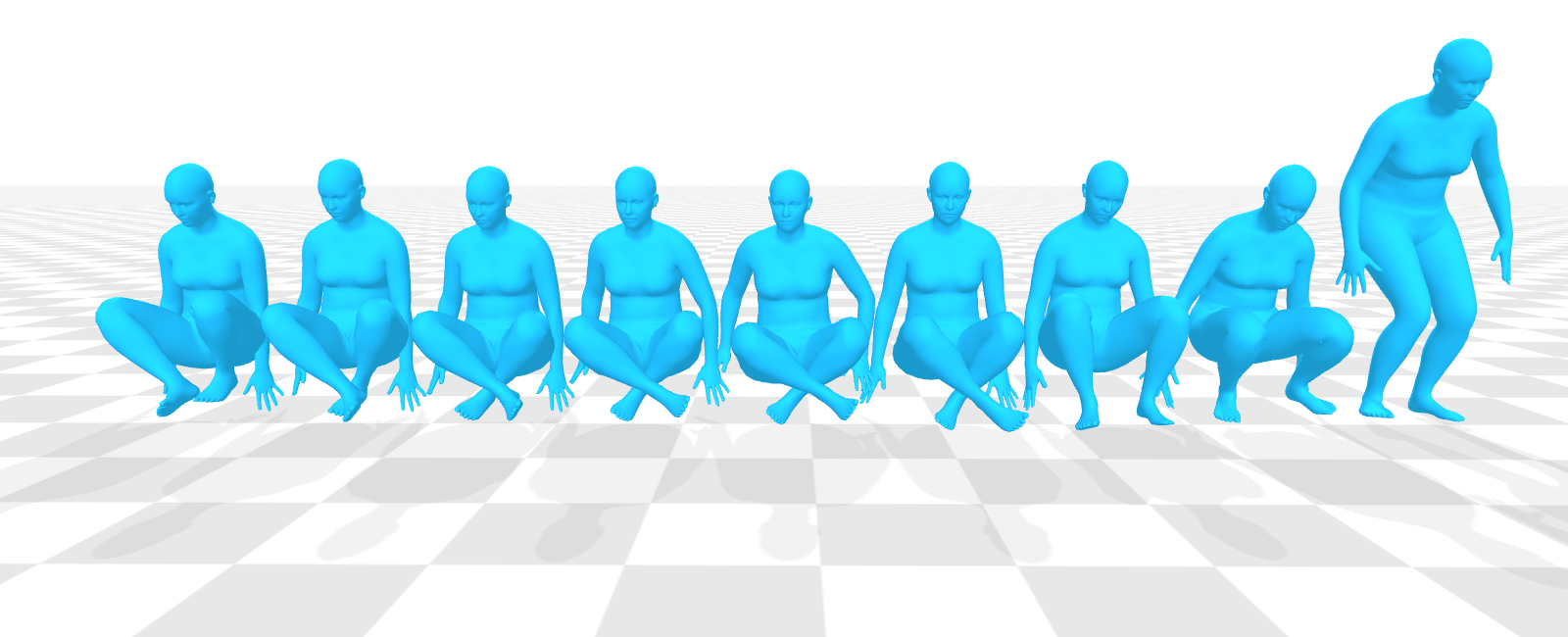} \\
\hline
\end{tabular}
\end{table*}

%在定性对比实验中，本方法与SimMotionEdit的比较结果显示，我们有效解决了SimMotionEdit在局部动作编辑中存在的显著问题。具体而言，在前两个示例中，SimMotionEdit未能准确响应“faster”、“slower”等速度相关指令，并出现了明显的肢体动作误编辑：在第一个示例中，本应编辑的右臂被错误地映射至左臂；第二个示例中，右臂则产生了预期之外的非自然动作。在编辑精度方面，第三个示例虽成功实现了反向动作生成，但对源动作的关键特征保持不足，还原度较低。而在第四和第五个示例中，本方法能够准确理解“是否站立”的语义意图，并正确生成对应动作，体现出在动作–文本语义对齐方面的显著优势。
% In the qualitative comparative experiments, the results of our method versus SimMotionEdit demonstrate that we have effectively addressed significant issues present in SimMotionEdit regarding local motion editing. Specifically, in the first two examples, SimMotionEdit failed to accurately respond to speed-related instructions such as ``faster'' and ``slower'' and exhibited noticeable erroneous edits in limb movements: in the first example, the right arm that should have been edited was incorrectly mapped to the left arm; in the second example, the right arm produced unexpected unnatural motions. In terms of editing precision, while the third example successfully achieved reverse motion generation, it inadequately preserved key features of the source motion, resulting in low fidelity. In the fourth and fifth examples, our method accurately understood the semantic intent of ``whether standing or not'' and correctly generated the corresponding motions, demonstrating a significant advantage in motion–text semantic alignment.
In qualitative comparative experiments, our method demonstrates clear advantages over SimMotionEdit in addressing key issues related to local motion editing. As shown in the first two examples, SimMotionEdit fails to accurately respond to speed-modification instructions such as faster or slower and introduces noticeable artifacts in limb motion: in the first case, the intended edit on the right arm was incorrectly applied to the left arm; in the second, the right arm exhibits unnatural movement. Regarding editing precision, although the third example successfully generates a reversed motion, it fails to preserve essential features of the source motion, leading to low fidelity. In the fourth and fifth examples, our method correctly interprets the semantic intent of \emph{whether standing or not} and generates corresponding motions accordingly, demonstrating a strong capability in motion–text semantic alignment.

\subsection{Ablation study} 
%为系统验证本方法中各核心模块的贡献，我们设计并进行了两组消融实验。由于部位感知调制模块（PMM） 与部位相似度监督机制（PSM） 在功能上紧密耦合——前者负责学习自适应部位权重，后者则为其提供明确的训练信号——我们首先对它们进行组合消融，以探究其协同作用。随后，我们进一步评估了双向语义交互模块（BMI） 在整个框架中的重要性。
% To systematically validate the contributions of each core module in our method, we designed and conducted two sets of ablation experiments. Given that the Motion Part-Aware  Modulation (PMM) and the Part-level similarity curve supervision mechanism (PSM) are functionally tightly coupled—the former learns adaptive part weights, while the latter provides explicit training signals for it—we first performed a combined ablation on them to investigate their synergistic effects. Subsequently, we further evaluated the importance of the Bidirectional Motion Interaction(BMI) within the overall framework.
%We designed and conducted two sets of ablation studies. 
The Part-aware Motion Modulation (PMM) module learns adaptive part-wise weights, and the Part-level similarity curve Supervision mechanism (PSM) supplies explicit supervisory signals for this process. We ablate them jointly to evaluate their effectiveness. 

\begin{table}[h]
\centering
\caption{We conducted a combined ablation study on the Motion Part-Aware  Modulation (PMM) module and the Part-level similarity curve supervision mechanism (PSM) to verify their necessity.}
\begin{tabular}{c|c|cccc}
\hline
 \multirow{2}{*}{PMM}& \multirow{2}{*}{PSM}& \multicolumn{4}{c}{Generated-to-Target (Batch)} \\
\cline{3-6}
 & & R@1↑ & R@2↑ & R@3↑ & AvgR↓ \\
\hline
$\times$  & $\times$  & 70.83 & 83.12 & 87.71 & 2.31 \\
$\checkmark$ & $\times$  & 70.83 & 83.33 & 88.12 & 2.29 \\
$\times$  & $\checkmark$ & 72.08 & 83.96 & 89.17 & 2.19 \\
$\checkmark$ & $\checkmark$ & \textbf{73.96} & \textbf{85.83} & \textbf{90.21} & \textbf{1.92} \\
\hline
\end{tabular}
\label{tab:ablation_batch1}
\end{table}

As shown in Table~\ref{tab:ablation_batch1}, our method with both PMM and PSM achieves the best accuracy: R@1/R@2/R@3 are 73.96/85.83/90.21 and AvgR is 1.92. It demonstrates that the explicit supervision of the PSM module and the adaptive modulation of the PMM module are complementary and indispensable for text-driven motion editing. AvgR increases to 2.19 immediately when the PMM is not used, due to the lack of explicit part-level supervision for learning accurate weight allocation. And AvgR increases to 2.29 when the PSM is not used, which shows that PSM improves all retrieval metrics by providing strong part similarity signals.

\begin{table}[h]
\centering
\caption{To validate the importance of the Bidirectional Motion Interaction (BMI) module, we removed it from the full model (Ours) to have the one without BMI (w/o BMI), and compared its motion quality (M-Score) with the ground truth (GT) as well.}
\begin{tabular}{c|ccccc}
\hline
\multirow{2}{*}{Methods}& \multicolumn{5}{c}{Generated-to-Target (Batch)} \\
\cline{2-6}
 & R@1↑ & R@2↑ & R@3↑ & AvgR↓ & M-Score↑ \\
\hline
GT & 100.0 & 100.0 & 100.0 & 1.00 & -4.081 \\
W/o BMI & 72.08 & 84.58 & 89.79 & 1.94 & -4.432 \\
Ours &  \textbf{73.96} &  \textbf{85.83} &  \textbf{90.21} &  \textbf{1.92} &  \textbf{-4.114} \\
\hline
\end{tabular}
\label{tab:ablation_batch2}
\end{table}

%结果（见表X）表明，移除BMI后，模型性能（-4.432）相较于完整模型（-4.114）出现明显下降。这一结果凸显了BMI的核心作用：它建立的文本与动作之间的双向细粒度对齐，是保证编辑后动作语义一致性与自然度的关键。缺乏这一深层交互，模型无法充分理解文本指令的意图，导致生成质量受损。最终，我们的完整模型在动作质量上最接近真实动作（GT），充分证明了整个框架的有效性。
% Table~\ref{tab:ablation_batch2} shows that after removing BMI, the model's performance (-4.432) significantly declined compared to the full model (-4.114). This outcome highlights the critical role of BMI: the bidirectional fine-grained alignment it establishes between text and motion is key to ensuring the semantic consistency and naturalness of the edited motions. Without this deep interaction, the model fails to fully comprehend the intent of the text instructions, leading to compromised generation quality. Ultimately, our full model achieves motion quality closest to the ground truth (GT), fully demonstrating the effectiveness of the entire framework.
% Table~\ref{tab:ablation_batch2} shows that removing the BMI leads to a noticeable performance drop, with the M‑Score deteriorating from –4.114 to –4.432. This result underscores the critical role of BMI in establishing fine‑grained bidirectional alignment between text and motion, which is essential for maintaining semantic consistency and naturalness in the edited motions. Without this deep interaction, the model fails to adequately capture the intent of the textual instructions, resulting in reduced generation quality. Overall, the full model yields motion quality that closely approximates the ground truth, confirming the effectiveness of the complete framework.
We further evaluated the importance of the Bidirectional Motion Interaction (BMI) module.
Table~\ref{tab:ablation_batch2} shows that removing the BMI module causes a clear performance drop (M‑Score drops from –4.114 to –4.432). 
It confirms BMI's critical role in establishing fine‑grained bidirectional text-motion alignment. 
The full model achieves motion quality closest to ground truth, validating the complete framework.
\section{Conclusion} 
\label{sec:conclusion}

We have presented \mname, a novel fine-grained text-driven framework for human motion editing.
%Its core contributions include three key components: a Motion Bidirectional Interaction Module (BMI), a Part-aware Motion Modulation (PMM), and a Part-level similarity-curve Supervision Mechanism (PSM). 
A Part-aware Motion Modulation (PMM) module is designed to automatically predict temporal part-wise weights with a multi-part semantic similarity curve supervision. And a Bidirectional Motion Interaction (BMI) module is further introduced to effectively improve the sematic alignment between text instructions and generated motions. Extensive comparisons and ablation studies on the MotionFix~\cite{athanasiou2024motionfix} show the effectiveness and advantages of our method.

\mname\ has certain limitations. For instance, its computation of part similarity relies on explicit geometric distance metrics, can not fully capture high-level semantic relationships. Besides addressing these limitations, we also plan to extend our method to more challenging scenarios, such as sequential editing tasks with a memory mechanism.

%For instance, its computation of part similarity relies on explicit geometric distance metrics, which may fail to fully capture high-level semantic relationships. Addressing these limitations represents one direction for future work. Additionally, we plan to extend our approach to more challenging scenarios, such as sequential editing tasks requiring a memory mechanism.
%explore adaptive part modeling and implicit alignment methods to support more flexible and intelligent motion editing.
% The part division relies on fixed joint groupings, losing the flexibility for dynamic part variations across different motions. And 
%the part similarity is computed using explicit geometric distance metrics, can hardly capture high-level semantic relationships. Future work will thus explore adaptive part modeling and implicit alignment methods to support more flexible and intelligent motion editing.

% WARNING: do not forget to delete the supplementary pages from your submission 
%\input{sec/X_suppl}

{
    \small
    \bibliographystyle{ieeenat_fullname}
    \bibliography{main}
}

\end{document}